\documentclass[letterpaper]{nature}

\usepackage[left=1in,right=1in,bottom=1.1in,top=1in]{geometry}
\usepackage{setspace}   
\usepackage{bbm}  
\usepackage{mathtools}  
\usepackage[algoruled]{algorithm2e}  
\usepackage[colorlinks,citecolor=blue,urlcolor=blue,linkcolor=blue]{hyperref}  
\usepackage[font=footnotesize,labelfont=bf]{caption}  
\usepackage{lineno}   
\usepackage{xurl}  

\makeatletter
\let\LN@align\align
\let\LN@endalign\endalign
\renewcommand{\align}{\linenomath\LN@align}
\renewcommand{\endalign}{\LN@endalign\endlinenomath}
\makeatother

\title{\Large 
    \begin{center}
        Accurate Measures of Vaccination and\\Concerns of Vaccine Holdouts from Web Search Logs
    \end{center}}

\author
{\begin{center}
\vspace{-5mm}
Serina Chang$^{1,*}$, Adam Fourney$^{2}$, Eric Horvitz$^{2,\dagger}$\\[5mm]
\footnotesize{$^{1}\;$Department of Computer Science, Stanford University} \\[1mm]
\footnotesize{$^{2}\;$Microsoft Research} \\[1mm]
\footnotesize{$^{*}\;$Research performed during an internship at Microsoft.} \\[1mm]
\footnotesize{$^{\dagger}\;$Corresponding author. Email: horvitz@microsoft.com} \\[1mm]
\end{center}
}

\begin{document}
\maketitle
\thispagestyle{empty}  %

\begin{singlespace}
\begin{abstract}
To design effective vaccine policies, policymakers need detailed data about who has been vaccinated, who is holding out, and why.
However, existing data in the US are insufficient: reported vaccination rates are often delayed or missing, and surveys of vaccine hesitancy are limited by high-level questions and self-report biases.
Here, we show how large-scale search engine logs and machine learning can be leveraged to fill these gaps and provide novel insights about vaccine intentions and behaviors.
First, we develop a \textit{vaccine intent classifier} that can accurately detect when a user is seeking the COVID-19 vaccine on search.
Our classifier demonstrates strong agreement with CDC vaccination rates, with correlations above 0.86, and estimates vaccine intent rates to the level of ZIP codes in real time, allowing us to pinpoint more granular trends in vaccine seeking across regions, demographics, and time.
To investigate vaccine hesitancy, we use our classifier to identify two groups, \textit{vaccine early adopters} and \textit{vaccine holdouts}.
We find that holdouts, compared to early adopters matched on covariates, are 69\% more likely to click on untrusted news sites.
Furthermore, we organize 25,000 vaccine-related URLs into a hierarchical ontology of vaccine concerns, and we find that holdouts are far more concerned about vaccine requirements, vaccine development and approval, and vaccine myths, and even within holdouts, concerns vary significantly across demographic groups.
Finally, we explore the temporal dynamics of vaccine concerns and vaccine seeking, and find that key indicators emerge when individuals convert from holding out to preparing to accept the vaccine.
\end{abstract}
\end{singlespace}
\clearpage
\pagenumbering{arabic}  %

\newcommand\syc[1]{\textcolor{blue}{SYC: #1}}
\newcommand\eh[1]{\textcolor{ForestGreen}{EH: #1}}
\newcommand\af[1]{\textcolor{purple}{AF: #1}} 
\newcommand\github{\url{https://github.com/microsoft/vaccine_search_study}}
\newcommand\vi[1]{\tilde{p}(v, #1)}
{\customspacing{1}
\section*{Introduction}
COVID-19 vaccines provide significant protection against severe cases of SARS-CoV-2,\cite{polack2020vaccine,bernal2021delta} yet a large portion of the United States remains unvaccinated.
Effective vaccine policies---for example, where to place vaccine sites,\cite{mehrab2021vaccine,weintraub2023desert} how to communicate about the vaccine,\cite{chou2020emotion,vergara2021trust} and how to design campaigns to reach unvaccinated populations\cite{dai2021nudges,rabb2022nudges,athey2023interventions}---rely on detailed data about who is seeking vaccination, who is holding out, and why.
However, existing data are insufficient.\cite{lafraniere2022data} 
Reported vaccination rates are frequently delayed,\cite{abutaleb2021cdc} missing at the county-level and below,\cite{tolbert2021kff} and missing essential demographic data.\cite{krieger2021demo,goldfarb2021statedemo}
Surveys provide a starting point for understanding vaccine hesitancy but are often limited by high-level questions,\cite{census2021pulse} small or biased samples,\cite{troiano2021hesitancy,bradley2021surveys} and self-reporting biases (e.g., recall or social desirability bias),\cite{althubaiti2016biases,sethsd2014racial} especially in sensitive contexts such as vaccination.\cite{jimenez2014selfreport}

We demonstrate how large-scale search engine logs and machine learning can be leveraged to fill these gaps, enabling fine-grained estimation of vaccine rates and discovering the concerns of vaccine holdouts from their search interests.
We use billions of anonymized search logs from Bing and introduce two computational resources to extract meaning from unlabeled queries and clicks.
First, we develop a \textit{vaccine intent classifier} to detect when a user is seeking the COVID-19 vaccine on search.
Our classifier achieves areas under the receiver operating characteristic curve (AUCs) above 0.90 on held-out vaccine intent labels in all states, and we observe strong agreement with CDC vaccination rates across states ($r=0.86$) and over time ($r=0.89$).
Using our classifier, we can estimate vaccine intent rates to the level of ZIP code tabulation areas (ZCTAs), approximately 10x the granularity of counties and preceding lags in reporting.
Our second key resource is a novel \textit{ontology of COVID-19 vaccine concerns} on search.
Our ontology consists of 25,000 vaccine-related URLs, clicked on by Bing users, that we organized into a hierarchy of vaccine concerns from eight top categories to 36 subcategories to 156 low-level topics.
Unlike surveys, our ontology discovers these concerns directly from users' expressed interests and explores them at multiple scales.
Furthermore, by measuring individuals' interest in each concern from their clicks, we capture revealed preferences, side-stepping potential biases in self-reporting.\cite{dumais2014log,sethsd2014racial}
We release our vaccine intent estimates and ontology of vaccine concerns online, along with our code.\footnote{Our data and code are available at \github.}

Combining our ontology with the vaccine intent classifier allows us to conduct a thorough analysis of how individuals' vaccine concerns relate to whether they decide to seek the vaccine.
We use our classifier to identify two groups of users---vaccine early adopters and vaccine holdouts---and compare their search behaviors.
We identify significant differences in their vaccine concerns and news consumption; for example, compared to early adopters matched on covariates, vaccine holdouts are 69\% more likely to click on untrusted news sites.
We find that vaccine concerns also differ significantly even within holdouts, varying across demographic groups.
Finally, we analyze the temporal dynamics of vaccine concerns and vaccine seeking, and discover that individuals exhibit telltale shifts in vaccine concerns when they eventually convert from holding out to preparing to accept the vaccine.

We use Bing search logs, which have been used to study other health issues such as shifts in needs and disparities in information access during the pandemic,\cite{suh2021population,suh2022info} 
health information needs in developing nations,\cite{abebe2019africa} 
experiences around cancer diagnoses,\cite{paul2015,paul2016} 
concerns rising during pregnancy,\cite{fourney2015pregnancy} 
nutritional patterns across the world,\cite{west2013} 
and medical anxieties associated with online search.\cite{white2009cyberchondria}
Our efforts build on prior work that extracts insights about the COVID-19 vaccine from digital traces, such as social media\cite{muric2021twitter,pierri2022misinfo,poddar2022twitter} and aggregated search trends,\cite{diaz2021fertility,bavadekar2021vsi,malahy2021insights} and other efforts to detect health conditions online, such as predicting depression from social media\cite{choudhury2013depression} and monitoring influenza from search queries.\cite{ginsberg2009flu}
Our work seeks to address the challenges of working with digital traces\cite{dumais2014log,olteanu2019social} and limitations of prior work\cite{ginsberg2009flu,lazer2014parable} by developing rigorous, human-in-the-loop methods to precisely detect user intents and interests (Sections~\ref{sec:methods-classifier} and~\ref{sec:methods-ontology}), correct for bias from non-uniform Bing coverage (Section~\ref{sec:methods-bing-coverage}), and validate our results against external data (Section~\ref{sec:methods-cdc}).
As one of the first works to use \textit{individual} search logs to study the COVID-19 vaccine, we have the rare opportunity to link vaccine outcomes (predicted by our classifier) to the same individual's search interests. 
While we report all results as aggregated over thousands of individuals, individual links reveal important dynamics between vaccine decision-making and vaccine concerns.
Our findings and resources can provide public health agencies and policymakers with finer-grained signals about vaccine seeking and holding out, helping to guide more effective, data-driven interventions.

\section*{Vaccine intent classifier}
We introduce a classifier to detect when users are expressing vaccine intent, i.e., seeking the COVID-19 vaccine on search. 
Vaccine intent can be expressed through unambiguous queries, such as [covid vaccine near me], which we detect using regular expressions that specify patterns to match in text.
However, vaccine intent can also be clarified through clicks on search results\cite{radlinski2010clicks}: for example, a user may issue an ambiguous query, such as [covid vaccine], then clarify their intent by clicking on the URL for the CVS COVID-19 vaccine registration page.
The challenge with URLs is that they are less formulaic than queries, so we cannot easily define regular expressions to identify URLs expressing vaccine intent.
Instead, we employ a series of graph-based machine learning techniques, combined with manual annotation, to identify URLs.

\begin{figure}[t]
    \centering
    \includegraphics[width=\linewidth]{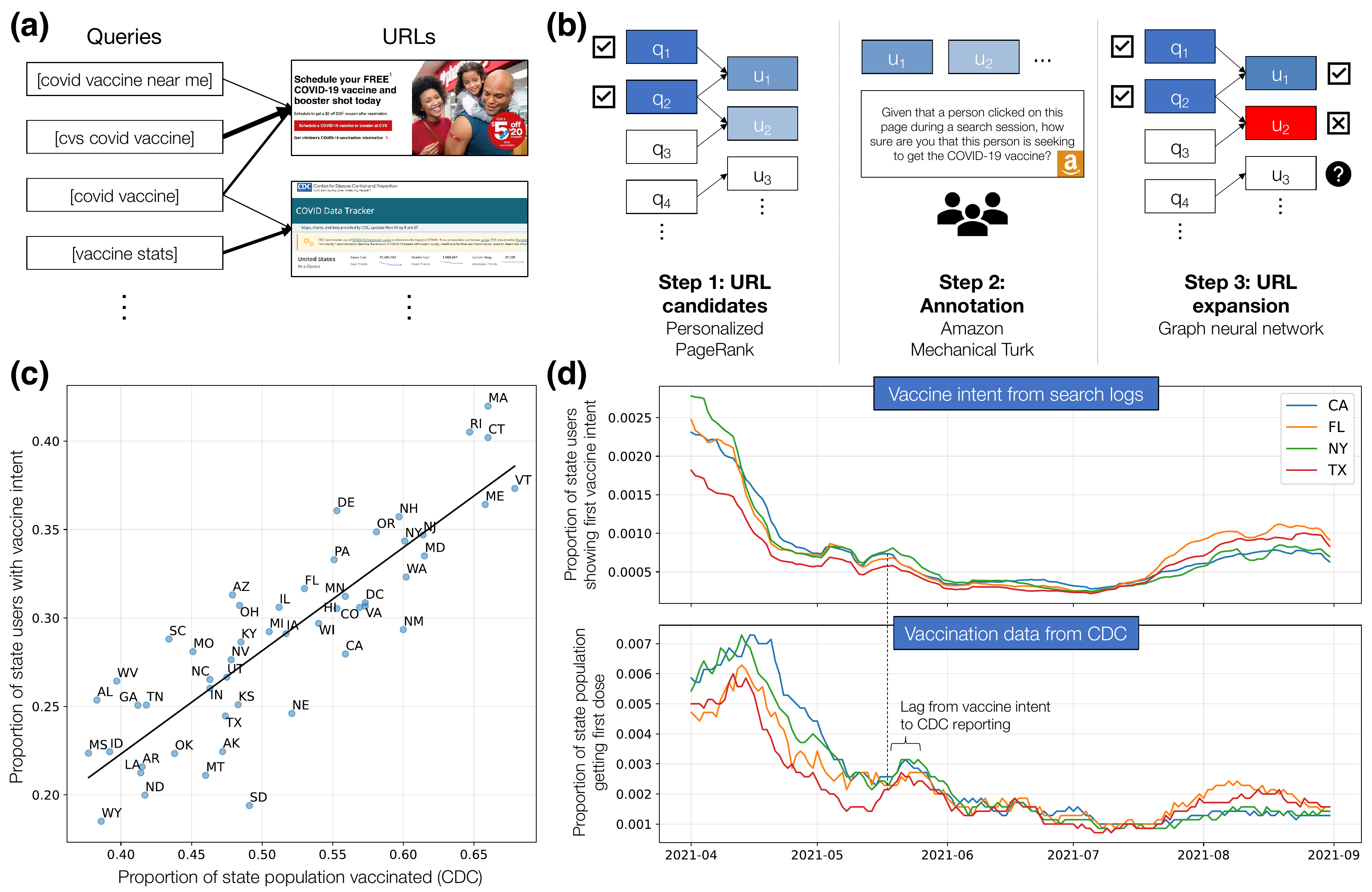}
    \caption{\textbf{Vaccine intent classifier}. 
    \textbf{(a)} Our computational approach centers on query-click graphs constructed from billions of Bing search logs.
    \textbf{(b)} Using these graphs, we introduce a three-step pipeline to identify vaccine intent URLs: generate URL candidates via Personalized PageRank; present URL candidates to annotators; and expand the final set of URLs with graph neural networks. Each step improves our coverage of users and correlation with CDC vaccination rates (Table \ref{tab:pipeline-results}).
    \textbf{(c)} Our vaccine intent estimates are highly correlated with state vaccination rates from the CDC. Here, we compare cumulative rates up to August 31, 2021 ($r=0.86$).
    \textbf{(d)} Our estimates are also highly correlated with CDC rates over time ($r=0.89$, median over states), with the CDC time series lagging by 7-15 days (IQR). Here, we visualize time series for the 4 largest states in the US, with extended results in Section \ref{sec:methods-cdc}.}
    \label{fig:classifier}
\end{figure}

\paragraph{Identifying vaccine intent URLs.}
Our key insight is that, while we cannot use regular expressions to identify URLs, we can use them to identify vaccine intent queries and then use those queries to identify URLs, based on common query-click patterns.
For example, vaccine intent queries such as [cvs covid vaccine] or [covid vaccine near me] may result in clicks on the CVS COVID-19  vaccine registration page.
To capture these patterns, we construct large-scale \textit{query-click graphs},\cite{craswell2007clickgraph,li2008click} which are bipartite networks between queries and URLs where an edge from a query to a URL indicates how often this query is followed by a click on this URL (Figure \ref{fig:classifier}a).
We use regular expressions to identify vaccine intent queries in the graph, and then propagate labels from these queries to URLs via Personalized PageRank (Figure \ref{fig:classifier}b, left).\cite{brin1998pagerank,kloumann2014pagerank}
This process enables us to identify URL candidates that likely express vaccine intent, without any URL labels.

We then present the URL candidates to annotators on Amazon Mechanical Turk and ask them to label whether these URLs indicate vaccine intent (Figure \ref{fig:classifier}b, middle).
We observe strong performance from our PageRank-based approach: even if positive labels require agreement from 3 annotators (out of 3-4), we find that 86\% of the URL candidates are labeled positive for vaccine intent (Figure \ref{fig:amt-vs-sppr}).
However, since manual annotation is expensive, we are only able to label around 2,000 URLs through this method.
To expand this set, we use these labels to train graph neural networks\cite{kipf2017gcn} (GNNs) to predict vaccine intent, so that we can use GNNs to predict labels for the remaining URLs (Figure \ref{fig:classifier}b, right).
Our GNNs demonstrate strong performance in all 50 states, with AUCs over 0.90 on held-out URLs labeled for vaccine intent (Figure \ref{fig:results-across-states}, Table \ref{tab:gnn-pretrain}).
Using our GNNs, we discover 11,400 more URLs that are highly indicative of vaccine intent.

\paragraph{Correcting for bias in vaccine intent estimates.}
We apply our classifier to Bing search logs from Feburary 1 to August 31, 2021 (Section \ref{sec:methods-data}) and identify 7.45 million active Bing users who have expressed vaccine intent through their queries or clicks.
However, before we can use the classifier to estimate regional rates of vaccine intent, we need to correct for potential sources of bias in our approach.
We decompose potential bias into two key sources (Section \ref{sec:methods-bias-decomp}): first, bias from non-uniform Bing coverage, and second, bias from non-uniform true and false positive rates of our classifier.
By correcting for non-uniform Bing coverage (Section \ref{sec:methods-bing-coverage}) and demonstrating that our classifier's true and false positive rates do not significantly differ across regions (Section \ref{sec:methods-bias-classifier}), our vaccine intent estimates should, theoretically, form unbiased estimates of true vaccination rates. 
Supporting this claim are our empirical results showing that our vaccine intent estimates agree strongly with CDC vaccination rates.
Furthermore, to evaluate the representativeness of Bing data, we compare search trends for vaccine intent queries between Google and Bing and find that, even before applying corrections to Bing data, the trends are highly correlated (Figures \ref{fig:google-time} and \ref{fig:google-states}).

\paragraph{Our vaccine intent estimates are highly correlated with CDC data.}
When we compare our vaccine intent estimates to state-level vaccination rates from the Centers for Disease Control and Prevention (CDC), we observe strong correlation ($r=0.86$) on cumulative rates at the end of August 2021 (Figure \ref{fig:classifier}c).
Notably, we find that the correlation drops to $r=0.79$ if we do not correct for Bing coverage in our estimates. 
If we only use queries to detect vaccine intent, the correlation drops to $r=0.62$ and we lose 57\% of the users we identified with our full classifier, demonstrating the value of including URLs (Table \ref{tab:pipeline-results}).
Additionally, we compare our vaccine intent estimates to the CDC's vaccination rates over time.
We observe strong correlations here as well, especially if we allow the CDC time series to lag behind the vaccine intent time series. 
With lags of 7-15 days (IQR), the median correlation over states reaches $r=0.89$; without a lag, the median correlation drops to $r=0.78$.
The CDC's lag demonstrates an advantage of our classifier, as it can detect vaccine seeking in real time without delays from reporting.

\begin{figure}[t]
    \centering
    \includegraphics[width=\linewidth]{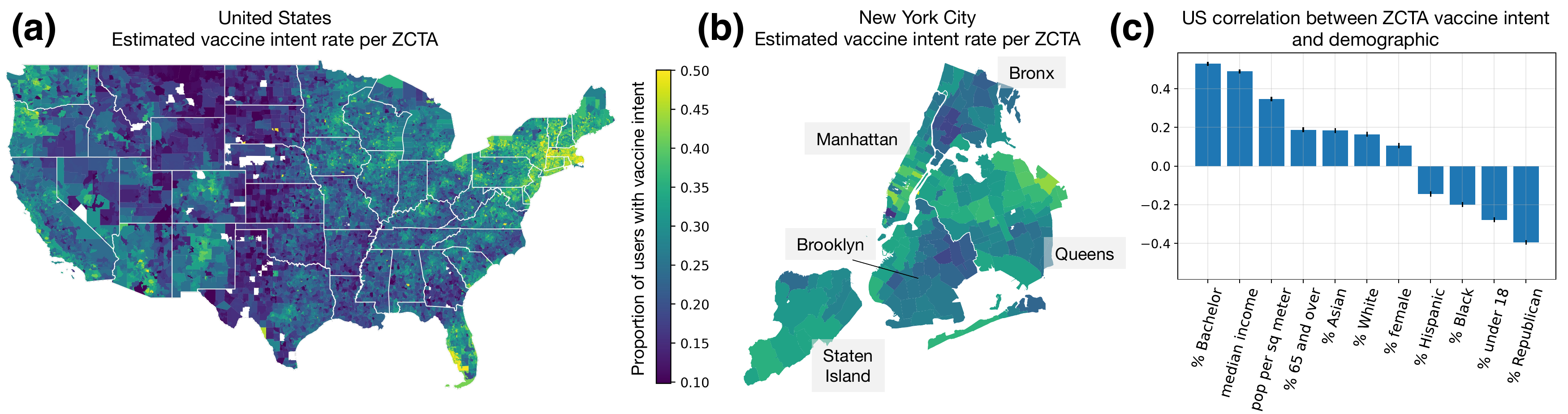}
    \caption{\textbf{Granular trends in vaccine seeking}. 
    \textbf{(a)} Using our classifier, we can estimate vaccine intent rates per ZCTA, approximately 10x the granularity of counties.
    \textbf{(b)} Zooming in on New York City shows that estimated vaccine intent rates vary substantially across ZCTAs, even within the same city or county.
    \textbf{(c)} We measure correlations between ZCTA vaccine intent rates and demographic variables to characterize demographic trends in vaccination.}
    \label{fig:vi-trends}
\end{figure}

\paragraph{Granular trends in vaccine seeking.}
Our vaccine intent classifier allows us to pinpoint who was seeking the COVID-19 vaccine, where, and when.
We estimate cumulative vaccine intent rates up to the end of August 2021 at the level of ZCTAs (Figure \ref{fig:vi-trends}a), approximately 10x the granularity of counties, which is the finest-grained vaccination data the CDC provides and, still, with many counties missing or having incomplete data.\cite{tolbert2021kff}
We observe substantial heterogeneity in vaccine intent at the ZCTA-level, even within the same states and counties.
For example, when we focus on New York City, we see that Manhattan and Queens have higher vaccine intent rates, and within Queens, ZCTAs in the northern half have higher rates (Figure \ref{fig:vi-trends}b), aligning with reported local vaccination rates in New York City.\cite{nyt2021nyc}
We can also use our estimates to characterize demographic trends in vaccination.
When we measure correlations between ZCTA vaccine intent rate and different demographic variables, we find that overall demographic trends from our estimates align closely with prior literature.\cite{kreps2020factors,troiano2021hesitancy,joshi2021predictors,yasmin2021review}
For example, we observe strong positive correlations with education, income, and population density, and a strong negative correlation with percent Republican (Figure \ref{fig:vi-trends}c).
However, we discover more nuanced trends when we look closer.
Demographic trends vary significantly across states (Figure \ref{fig:state-demo}), especially for race and ethnicity, and trends change over time (Figure \ref{fig:demo-quartile}).
For example, we estimate that older ZCTAs were much likelier to seek the vaccine early in 2021 but this trend fell over time, 
reflecting how the US vaccine rollout initially prioritized seniors,\cite{cnbc2021seniors}
and we see an increase in vaccine intent from more Republican ZCTAs in summer 2021,
reflecting new calls from Republican leaders to get vaccinated\cite{wapo2021gop} and a self-reported uptick in vaccinations among Republicans.\cite{gallup2021sept}
Thus, our classifier both confirms existing findings and enables new analyses with finer granularity across regions, demographics, and time.

\section*{Search concerns of vaccine holdouts}
We use our vaccine intent classifier to identify two groups: \textit{vaccine early adopters}, who expressed their first vaccine intent before May 2021, and \textit{vaccine holdouts}, who waited until July 2021 to show their first vaccine intent, despite becoming eligible by April.\footnote{We did not consider as holdouts those who never showed vaccine intent during our study period, since those users may have gotten their vaccine in ways that are not visible via search data, e.g., a walk-in appointment. In comparison, individuals who did not show their first vaccine intent until July 2021 likely did not receive the vaccine before.}
Comparing the search interests of these two groups allows us to discover relationships between expressed vaccine concerns, news consumption, and vaccine decision-making.
To reduce potential confounding, we match each holdout with a unique early adopter from the same county and with a similar average query count, since we know that the populations seeking vaccination changed over time and we do not want our comparisons to be overpowered by regional or demographic differences. In our following analyses, we compare the search interests of the matched sets, with over 200,000 users in each set.

\paragraph{Vaccine holdouts are more likely to consume untrusted news.}
First, we analyze the trustworthiness of news sites clicked on by vaccine holdouts versus early adopters.
We use ratings from Newsguard, which assigns trust scores to news sites based on criteria such as how often the site publishes false content and how it handles the difference between news and opinion.\cite{newsguard}
We find that, in the period while vaccine holdouts were eligible but still holding out (April to June 2021), holdouts were 69\% (95\% CI, 67\%-70\%) likelier than their matched early adopters to click on untrusted news, defined by Newsguard as domains with trust scores below 60.
Furthermore, we see that as the trust score from Newsguard degrades, the likelier it was that holdouts clicked on the site, relative to early adopters (Figure \ref{fig:concerns}a).
For example, sites that are known for spreading COVID-19 misinformation, such as infowars.com,\cite{nyt2020jones} RT.com,\cite{nyt2021rt} and mercola.com\cite{nyt2021mercola}, were much likelier to be clicked on by holdouts.
These results extend prior work linking vaccine hesitancy and misinformation\cite{muric2021twitter,pierri2022misinfo,loomba2021misinfo} by directly analyzing the online news consumption of vaccine holdouts and quantifying their heightened consumption of misinformation.

\begin{figure}
    \centering
    \includegraphics[width=\linewidth]{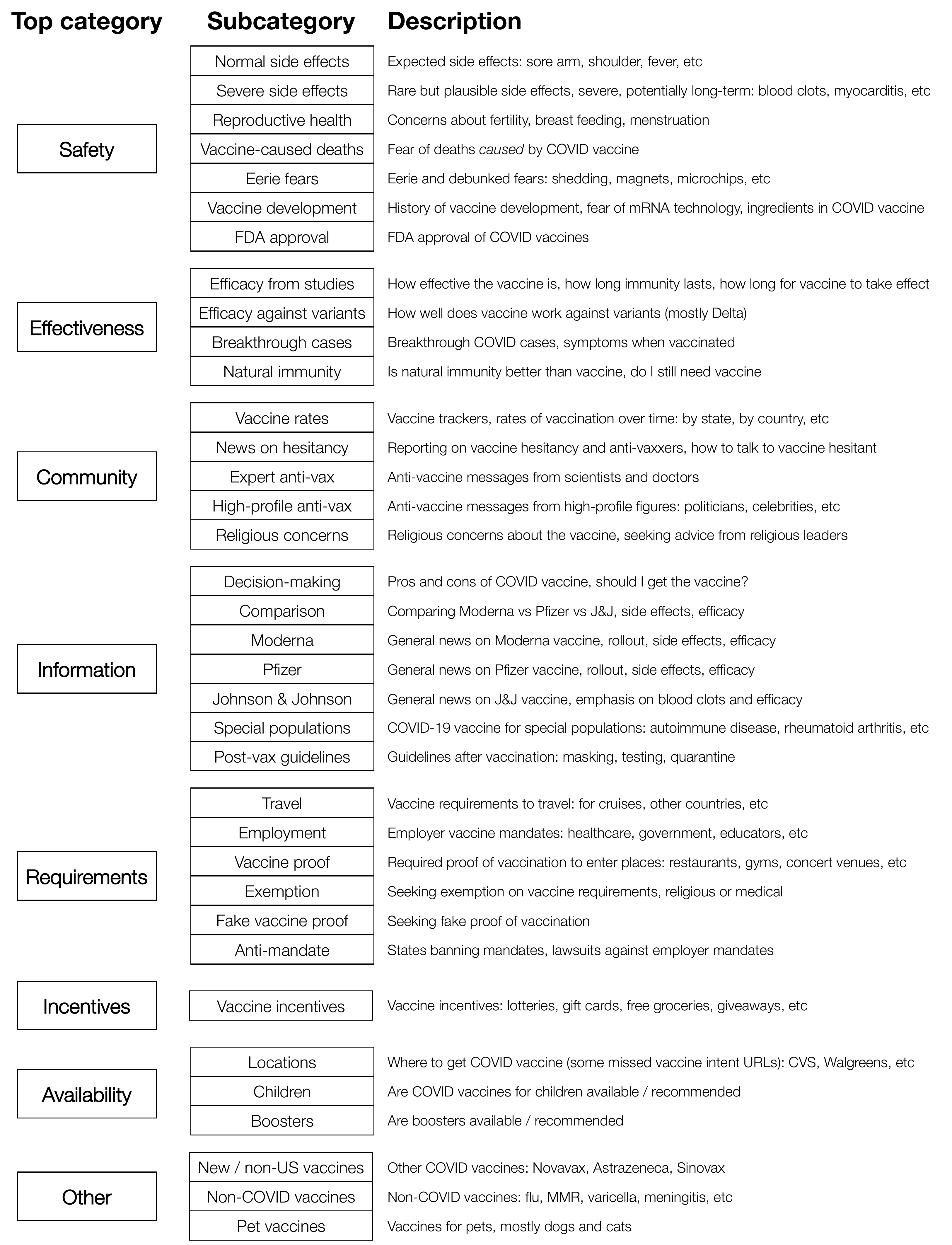}
    \caption{Our ontology of vaccine concerns consists of 8 top categories and 36 subcategories.}
    \label{fig:ontology}
\end{figure}
\paragraph{Ontology of vaccine concerns on search.}
To characterize vaccine-related search interests in far more detail, we construct a hierarchical ontology of vaccine concerns, defined in terms of 25,000 vaccine-related URLs that were clicked on by early adopters or holdouts.
First, using the Louvain algorithm for community detection on graphs,\cite{blondel2008louvain} we automatically partition the URLs into 156 clusters (each containing around 100-500 URLs).
Based on these clusters, which are remarkably coherent (Table~\ref{tab:cluster-samples}), we design a comprehensive set of subcategories and top categories, and sort the clusters accordingly.
For example, we identify one cluster of news stories announcing vaccine passport requirements in cities such as New York City and Los Angeles,\cite{cnn2021nyc,deadline2021la} which we sort under the proof of vaccination subcategory and Vaccine Requirements top category.
This bottom-up approach allows us to discover and measure vaccine concerns directly from users' search interests and analyze them at multiple scales, providing complementary insights to more traditional surveys.

In Figure~\ref{fig:ontology}, we summarize our resulting ontology, which consists of 8 top categories and 36 subcategories.
Some top categories encompass a number of distinct subcategories: for example, under Vaccine Safety, we include normal side effects, severe side effects, concerns about reproductive health, vaccine history and development, FDA approval, fear of vaccine-caused deaths, and ``eerie'' fears (e.g., myths about vaccine shedding or becoming magnetic\cite{cdc2023myths}).
At the top category-level, we find that vaccine holdouts are, by far, the most concerned about Vaccine Safety, which accounts for 23\% of their vaccine-related clicks, followed by Vaccine Information (10\%) and Vaccine Requirements (9\%).
We also observe changes in interests over time (Figure \ref{fig:concerns}b): for example, interest in Vaccine Incentives increased in May 2021, when incentives were introduced,\cite{usnews2021incentives} and interest in Vaccine Effectiveness grew in June, following the spread of the Delta variant.

\begin{figure}[t]
    \centering
    \includegraphics[width=\linewidth]{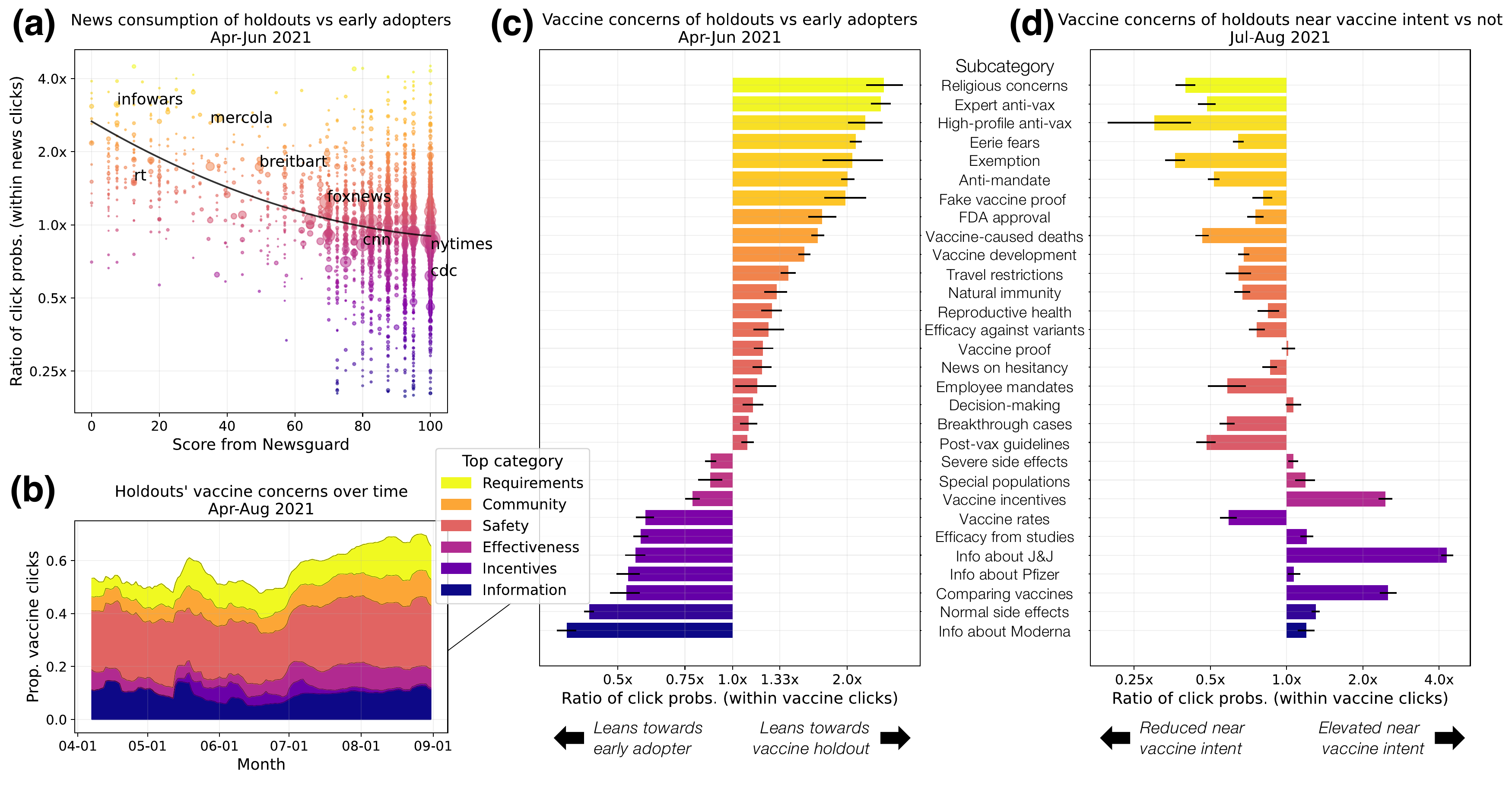}
    \caption{\textbf{Vaccine concerns and news consumption}. 
    In all subfigures, news/categories are colored from yellow to dark purple to represent most holdout-leaning to most early adopter-leaning.
    \textbf{(a)} The lower the trust rating from Newsguard, the likelier it is that vaccine holdouts click on the news site, relative to early adopters.
    \textbf{(b)} Holdouts' top category concerns include Vaccine Safety, Requirements, and Information, with varying proportions over time. 
    \textbf{(c)} Comparing holdouts vs. early adopters' relative probabilities of clicking on each subcategory (from April to June 2021) reveals each group's distinctive concerns. 
    \textbf{(d)} Near when holdouts express vaccine intent ($\pm$3 days) in July and August 2021, their concerns become much more like the concerns of early adopters, with a few important differences.}
    \label{fig:concerns}
\end{figure}

\paragraph{Distinctive concerns of holdouts vs. early adopters.} 
Our ontology allows us to compare the vaccine concerns of holdouts and their matched early adopters.
First, during the period from April to June 2021, we find that holdouts were 48\% less likely than early adopters to click on any vaccine-related URL.
Furthermore, their distribution of concerns within their vaccine-related clicks differed significantly (Figure \ref{fig:concerns}c, Table \ref{tab:holdout-ea-ratios}).
Using the subcategories from our ontology, we find that 
holdouts were far more interested in religious concerns about the vaccine; anti-vaccine messages from experts and high-profile figures; avoiding vaccine requirements by seeking exemptions, banning mandates, or obtaining fake proof of vaccination; eerie fears and vaccine-caused deaths; and FDA approval and vaccine development.
In comparison, early adopters were much more concerned about normal side effects, vaccine efficacy, comparing different types of vaccines, and information about each vaccine (Moderna, Pfizer, and Johnson \& Johnson).
These differences reveal the importance of a fine-grained ontology; for example, at the top category level, we would see that both groups were interested in Vaccine Safety but miss that early adopters were more concerned about normal and severe side effects, while holdouts were more concerned about eerie fears and vaccine-caused deaths.
Our approach also allows us to study \textit{who} is expressing these concerns in greater granularity.
Even within holdouts, we observe significant variability in concerns across demographic groups (Figure \ref{fig:concerns-demo}).
For example, holdouts from more Democrat-leaning ZCTAs were particularly concerned about FDA approval and vaccine requirements, while holdouts from more Republican-leaning ZCTAs were more concerned about eerie fears and vaccine incentives.

\paragraph{Holdouts appear like early adopters when seeking the vaccine.}
In our final analysis, we exploit the fact that all of our vaccine holdouts eventually expressed vaccine intent to explore how vaccine concerns change as an individual converts from holdout to adopter.
From July to August 2021, we analyze how holdouts' vaccine concerns change in the small window ($\pm 3$ days) surrounding their expressed vaccine intent, compared to their typical concerns outside of that window.
We find that in those windows, holdouts' vaccine concerns nearly reverse, such that they look much more like early adopters than their typical selves (Figure \ref{fig:concerns}d nearly reverses \ref{fig:concerns}c).
During this time, holdouts become far more interested in the Johnson \& Johnson vaccine, comparing different vaccines, and vaccine incentives, and less interested in anti-vaccine messages and vaccine fears (Table \ref{tab:near-vi-ratios}).
Notably, not all early adopter-leaning concerns reverse as dramatically; for example, even while expressing vaccine intent, holdouts remain less interested in the Pfizer and Moderna vaccines, which may reflect how vaccine hesitant individuals were quicker to accept the one-shot Johnson \& Johnson vaccine, instead of the two-shot mRNA vaccines.\cite{nyt2021jj,healthline2021jj}
Furthermore, there are some early adopter-leaning concerns that holdouts do not pick up on during this time, such as interest in vaccine rates. We hypothesize that these concerns are more reflective of an early adopter ``persona'' rather than of concerns that would become relevant when seeking the vaccine, such as comparing different vaccines or vaccine incentives.
\section*{Discussion}
We have demonstrated how large-scale search logs and machine learning can be leveraged for fine-grained, real-time monitoring of vaccine intent rates and identification of individuals' concerns about vaccines.
There are limitations to our approach: for example, while we can achieve finer granularity than existing data, we still miss within-ZCTA heterogeneity in vaccine intent.
Furthermore, our efforts to minimize bias in our estimates are substantial but imperfect (e.g., we can only approximate the true and false positive rates of our classifier).
We also assume in this work that vaccine intent can be detected through single queries or clicks, but more sophisticated models could incorporate entire search sessions or browsing data beyond search.
However, in favor of simplicity and considerations of privacy, we label vaccine intent at the query and click-level.
We are mindful throughout this work of the need to balance privacy and social benefits when using potentially sensitive user data.
For this reason, we only link as much as we need at the individual-level (e.g., without demographic attributes or browsing data) and only report aggregated results.

Despite these limitations, our resources demonstrate strong agreement with existing data and enable analyses that have not been available before.
Our vaccine intent classifier achieves high correlations with vaccination rates reported by the CDC, but it also allows us to estimate vaccine rates down to the ZCTA-level in real time.
This spatial granularity supports more precise analyses and interventions; for example, the finer-grained estimates can help public health officials to identify under-vaccinated communities, informing where to place vaccine sites or whom to prioritize in online or real-world outreach programs.
The real-time nature of our vaccine intent signals opens up opportunities to more effectively assess the effects of interventions, such as
determining if a new public service announcement is responsible for a proximal rise in vaccine interest, or understanding how people learned about a new vaccination location through search.

While our vaccine intent classifier can be harnessed to provide insights about where to intervene and for whom, our novel ontology and analyses of vaccine concerns inform \textit{how} to intervene. 
Search logs offer a glimpse into individuals' genuine interests and exposure to content, without depending on self-reported survey data or user-generated content on social media. 
By examining holdouts' online news consumption and specific vaccine concerns, and observing how these concerns change over time and across demographics, our findings could help shape messaging strategies in vaccination campaigns.
Lastly, our observation that holdouts resemble early adopters when they eventually seek vaccination indicates that individuals might follow similar paths towards vaccine acceptance. Future work could model these trajectories, try to identify key influences (e.g., vaccine mandates), incentives (e.g., lotteries), or assurances (e.g., trusted content on the rarity of severe side effects) and use these models to allocate limited resources for interventions. 

We focus in our research on the COVID-19 pandemic in the US. Our methods demonstrate the potential of large-scale computing platforms, utilized by millions nationwide, to offer valuable anonymized public health signals that might be otherwise challenging to acquire. The US Constitution places public health responsibilities in state governments. Thus, public health programs in the US largely depend on a decentralized, diverse patchwork of state and county public health agencies. In the context of the current state of affairs with the delivery of public health services in the US, inferences and insights based on data drawn from continent-spanning computing platforms methods can play a valuable integrative role. We also see value of the methods in countries with more centralized approaches to public health; we hope that future work can extend the methods we have developed to study vaccination behaviors in other languages and countries, as well as examining the deployment and adoption of additional vaccines.

To facilitate policy impact and future research, we have released our vaccine intent estimates and our ontology of vaccine concerns.
We hope that these resources will be useful to policymakers and researchers so that they can conduct detailed analyses of COVID-19 vaccine rates, such as evaluating the efficacy of vaccine distribution plans and studying disparities in vaccination rates.
The ontology can also be employed widely in web and social media research. For example, it can be used to study how certain classes of URLs (e.g., eerie fears) are disseminated on social media or surfaced by search engines.
Finally, we note that our graph-based machine learning techniques for intent detection on search are applicable beyond vaccines, and could be applied to precisely detect other intents of interest, such as registering to vote or filing for stimulus checks.
More broadly, we hope that our work can serve as a roadmap for researchers of how to derive rigorous behavioral and health insights from search logs, including how to precisely detect user intents and interests, evaluate and correct for bias in estimates, validate predictions against external data, and release resources to promote reproducibility, transparency, and future work.

}

\clearpage
{\customspacing{1.2}
\renewcommand\thesection{M\arabic{section}}
\setcounter{section}{0}

\section*{Methods}
The Methods section is structured as follows:
In Section \ref{sec:methods-data}, we discuss the datasets we use, including Bing search logs, CDC vaccination rates, and data from the US Census.
In Section \ref{sec:methods-classifier}, we describe our methods to develop a vaccine intent classifier. 
In Section \ref{sec:methods-bias}, we discuss how we apply our classifier to estimate regional vaccine intent rates, correct for and evaluate bias in our estimates, and compare to external data from CDC and Google.
In Section \ref{sec:methods-ontology}, we describe our methods to construct our ontology of vaccine concerns.
Finally, in Section \ref{sec:methods-analysis}, we provide details about our main analyses and describe supplemental analyses.

\section{Datasets} \label{sec:methods-data}
\paragraph{Bing search logs.} 
Our work leverages billions of anonymized search logs from Bing.
Bing is the second largest search engine worldwide and in the US, with a US market share of around 6\% on all platforms and around 11\% on desktop.\cite{search2021market}
Despite having non-uniform coverage across the US, Bing has enough penetration across the country that we can estimate representative samples after applying inverse proportional weighting (Section \ref{sec:methods-bing-coverage}).
The Bing data we use consist of individual queries made by users, where for each query, we have information including the text of the query, an anonymized ID of the user, the timestamp, the estimated geolocation (ZIP code, county, and state), and the set of URLs clicked on, if any.
Since our work is motivated by insufficient vaccine data and vaccine concerns in the US, we limit our study to search logs in the US market.
However, the methods we introduce could be extended to study vaccination rates and vaccine concerns in other languages and countries.
We apply our vaccine intent classifier (Section \ref{sec:methods-classifier}) to Bing search logs from February 1 to August 31, 2021.
February 2021 was the earliest that we could study following data protection guidelines, which allow us to store and analyze search logs up to 18 months in the past.
We end in August 2021, since the FDA approved booster shots in September and our method is not designed to disambiguate between vaccine seeking for the primary series versus boosters.
Our work was approved by the Bing product team, in addition to other privacy officers at Microsoft (\textbf{Data ethics}).

\paragraph{CDC vaccination rates.}
To evaluate our vaccine intent classifier, we compare it to vaccination rates reported by the Centers for Disease Control and Prevention (CDC) (Section \ref{sec:methods-cdc}). The CDC provides daily cumulative vaccination rates at the levels of states\cite{cdc_state_vax} and counties.\cite{cdc_county_vax}
They provide different measures, including the total number and percentage of population who have received at least one dose, completed a primary series, received a booster shot, received a second booster shot, and so on.
CDC data are essential but limited, with a substantial portion of county-level data missing.
These limitations are one of the motivations of our work, since we hope that our vaccine intent classifier can serve as a complementary resource to monitor vaccination rates, especially for smaller regions that the CDC does not cover. 

\paragraph{US Census.}
We estimate vaccine intent rates at the level of ZIP code tabulation areas (ZCTAs)\cite{census_zctas}, since they are the smallest Census-tracked unit to which we can reliably map Bing queries and users (Section \ref{sec:methods-bing-coverage}).
To characterize demographic trends in vaccine intent, we use ZCTA data from the US Census' 2020 5-year American Community Survey.\cite{census_acs}
The demographic variables we use are total population size, percent female, percent of different age groups (e.g., under 18, over 65), percent of different races/ethnicities (White, Black, Asian, Hispanic), percent with Bachelor degree or higher, and population per square meter (in log scale), which divides the population size by the ZCTA's land area.\cite{zcta2020layout}
To create map visualizations (e.g., Figure \ref{fig:vi-trends}a), we also use the 2020 ZCTA, county, and state shapefiles provided by the US Census.\cite{census_shapefiles}

\paragraph{Elections data.}
To capture political lean per region, we use county-level data from the 2020 US presidential election, which we purchased online.\cite{election_data}
In our analyses, we use ``percent Republican'' as a variable, i.e., the percentage of overall votes cast in the county that went to the Republican nominee Donald Trump.

\paragraph{Newsguard data.}
We use data from Newsguard to label the trustworthiness of different news sites.
Newsguard assigns numerical trust scores to news sites based on nine journalistic criteria, such as how often the site publishes false content, how responsibly it collects and presents information, how it handles the difference between news and opinion, and how transparent it is about its funding.\cite{newsguard}
News sites with scores above 60 are considered Trusted and below 60 are considered Untrusted.
They also categorize other sites as Satire or Platform, but these sites are not given numerical scores since the criteria are not as relevant.
In our analysis, we focus on the Trusted and Untrusted sites with numerical scores (Section \ref{sec:methods-concerns}).

\paragraph{Google search trends.}
To evaluate the representativeness of Bing search trends, we compare them to Google search trends (Section \ref{sec:methods-google}). 
Google allows individuals to view aggregated, normalized search trends for any query with enough users.\cite{google_trends}
The trends for that query over time and across subregions (e.g., US states) are then available for download.

\paragraph{Data ethics.} 
Our work was approved by both the Microsoft IRB office, and by an internal privacy review process which included privacy officers from both Microsoft Research and the Bing product team. Together, we worked to ensure that our use of Bing search logs was consistent with Bing's privacy policy (which explicitly lists research as a possible use of search data), and with relevant company best-practices.
When we use search logs, we are mindful of the need to balance privacy and social benefits when using potentially sensitive user data. 
While we study individual search logs, since we need to be able to link individual vaccine outcomes (as predicted by our classifier) to search interests, those sessions are assembled using only anonymous user identifiers, which are disassociated from any specific user accounts or user profiles, and cannot be linked to any other Microsoft products. 
Likewise, in this anonymous view of the logs, location and demographic data were limited to ZIP code-level accuracy. Finally, we are careful to only report results aggregated over thousands of individuals.
Aside from Bing search logs, all of the data sources we use here are publicly available and aggregated over many individuals.

\section{Vaccine intent classifier} \label{sec:methods-classifier}
We develop a vaccine intent classifier to detect users who are expressing vaccine intent, i.e., seeking the COVID-19 vaccine with web search.
As we discuss in the main text, we can use regular expressions to identify queries expressing vaccine intent, such as [where can i get a covid vaccine].
However, intent can also be clarified through clicks on search results,\cite{radlinski2010clicks} such as clicking on the CVS registration page for the COVID-19 vaccine. 
The challenge with URLs is that we cannot easily define regular expressions to identify the ones expressing vaccine intent.
Instead, our approach is to construct query-click graphs, then to use a combination of graph-based machine learning techniques and manual annotation to identify a large set of vaccine intent URLs.
Our method consists of three steps: use personalized PageRank to propagate labels from queries to URLs, so that we can generate a set of URL candidates for manual annotation (Section \ref{sec:methods-sppr}); 
present the URL candidates to external annotators on Amazon Mechanical Turk to label as vaccine intent or not (Section \ref{sec:methods-amt});
use the labels from the previous step to train graph neural networks so that we can further expand our set of vaccine intent URLs (Section \ref{sec:methods-gnn}).

\subsection{Personalized PageRank for URL candidates} \label{sec:methods-sppr}
\paragraph{Constructing query-click graphs.}
Our first goal is to construct query-click graphs that broadly cover searches related to COVID-19 and the vaccine.
First, we collect all queries that contain any word from a long list of keywords, including ``covid'', ``pandemic'', ``vaccin'', ``cdc'', ``moderna'', and so on.\footnote{The full list is \{``covid'', ``corona'', ``pandemic'', ``cov19'', ``virus'', ``variant'', ``vaccin'', ``vacin'', ``vax'', ``dose'', ``shot'', ``booster'', ``rollout'', ``roll out'', ``fda'', ``cdc'', ``johnson'', ``jj'', ``janssen'', ``pfizer'', ``phizer'', ``biontech'', ``moderna'', ``astrazeneca'', ``mrna''\}. We constructed this list by starting with a smaller core set, \{``covid'', ``vaccine'', ``vaccin'', ``booster''\}, collected queries that included any of those words, and looked through the top 1,000 words that appeared in those queries.}
Then, we also collect all queries and clicks that co-occurred in a search session with any of these queries.
Using co-occurrence allows us to capture vaccine-related queries and URLs that do not include the keywords, such as California's vaccine scheduling page, myturn.ca.gov.
We construct a query-click graph from all of these queries and clicks, with queries and URLs as nodes. 
Our graph consists of two types of edges: first, an edge from query A to query B represents that A preceded B in a search session; second, an edge from a query to a URL represents that searching that query led to a click on that URL.
In both cases, the edge weight indicates the number of times this relationship appears in our data.

Since URLs may appear or disappear over time, we collect queries and clicks from two spread-out months in our study period, April 1-30, 2021 and August 1-31, 2021. 
We construct query-click graphs \textit{separately} for every US state, since we find that the classifier performs better across states when we build different graphs and models per state (Section \ref{sec:methods-bias-classifier}).
We also perform minor preprocessing at this step: we lower-case queries, drop queries that are implausibly long (over 100 characters), and drop clicks that are not on URLs (e.g., ``javascript:void'').

\paragraph{Personalized PageRank from query seed set (S-PPR).}
We define vaccine intent queries as those that are \textit{unambiguously} seeking the COVID-19 vaccine with web search. 
To be included, the query must include both a COVID-19 term (``covid'' or ``coronavirus'') \textit{and} a vaccine term (``vaccin'', ``vacin'', ``vax'', ``dose'', ``shot'', ``booster'', ``johnson'', ``pfizer'', or ``moderna''). 
In addition, the query must satisfy at least one of the following criteria: (1) matching some variant of ``find me a COVID-19 vaccine'', (2) containing appointment-related words (e.g., ``appointment'', ``sign up'') or location-seeking words (e.g., ``near me'', ``where can i get''), (3) containing a pharmacy name.
We try to capture a representative list of pharmacy names by including almost all pharmacies\footnote{Based on an inspection of the queries containing each pharmacy name, we drop a few names that are ambiguous (e.g., United is a pharmacy, but ``united'' can be confused with United States or United Airlines).} that provided the COVID-19 vaccine through the Federal Retail Pharmacy Program for COVID-19 Vaccination, which includes 21 pharmacy partners and their many subsidiaries.\cite{cdc_pharmacies}

To identify URL candidates for vaccine intent, we use personalized PageRank.
Personalized PageRank\cite{brin1998pagerank} is a common technique for seed expansion, where a set of seed nodes in a graph are identified as members of a community, and one wishes to expand from that set to identify more community members.\cite{kloumann2014pagerank}
In our case, the vaccine intent queries act as our seed set, and our goal is to spread the influence from the seed set over the rest of the query-click graph.
For a given seed set $S$, personalized PageRank derives a score for each node in the graph that represents the probability of landing on that node when running random walks from $S$.
The hyperparameter $\alpha$ controls the lengths of the random walks by defining the probability of continuing the random walk versus teleporting back to the seed set.
In our work, we use the default $\alpha=0.85$.
Personalized PageRank naturally trades off between favoring nodes that are closer to the seed set (if random walks are shorter) and nodes that are central in the network (if random walks are longer).
These are also the two desiderata of the URL candidates we hope to find: they should be close to the vaccine intent queries in the graph and, to achieve high utility from labeling, they should be central and high-degree (i.e., we would not want to ``spend'' a label on a URL that is rarely clicked on).

Thus, we run personalized PageRank from the seed set of vaccine intent queries (S-PPR) to derive scores for all URLs in each query-click graph.
S-PPR also provided scores for all queries in the graph, but we found that our seed set was quite comprehensive in identifying unambiguous queries.
The top-ranked queries that were not in the seed set tended to be location-specific, such as [covid vaccine new york], which were suggestive of vaccine intent, but we decided were not unambiguous enough (e.g., they could be seeking vaccine locations in New York, but could also be seeking information about vaccine eligibility or requirements in New York).

\paragraph{Selecting URL candidates.}
\begin{table}
    \small
    \centering
    \begin{tabular}{c|c|c}
        State & Rank & URL \\
        \hline
        CA & 0 & \url{https://myturn.ca.gov/} \\
         & 1 & \url{https://www.cvs.com/immunizations/covid-19-vaccine} \\
         & 2 & \url{https://www.goodrx.com/covid-19/walgreens} \\
         & 3 & \url{https://www.costco.com/covid-vaccine.html} \\
         & 4 & \url{https://www.walgreens.com/topic/promotion/covid-vaccine.jsp} \\
         \hline
         NY & 0 & \url{https://covid19vaccine.health.ny.gov/} \\
         & 1 & \url{https://www.cvs.com/immunizations/covid-19-vaccine} \\
         & 2 & \url{https://www.walgreens.com/topic/promotion/covid-vaccine.jsp} \\
         & 3 & \url{https://vaccinefinder.nyc.gov/} \\
         & 4 & \url{https://www.goodrx.com/covid-19/walgreens} \\
         \hline
         TX & 0 & \url{https://www.cvs.com/immunizations/covid-19-vaccine} \\
         & 1 & \url{https://vaccine.heb.com/} \\
         & 2 & \url{https://www.walgreens.com/topic/promotion/covid-vaccine.jsp} \\
         & 3 & \url{https://corporate.walmart.com/covid-vaccine} \\
         & 4 & \url{https://dshs.texas.gov/covidvaccine/} \\
         \hline
         FL & 0 & \url{https://www.publix.com/covid-vaccine} \\
         & 1 & \url{https://www.cvs.com/immunizations/covid-19-vaccine} \\
         & 2 & \url{https://www.walgreens.com/topic/promotion/covid-vaccine.jsp} \\
         & 3 & \url{https://floridahealthcovid19.gov/vaccines/} \\
         & 4 & \url{https://www.goodrx.com/covid-19/walgreens  } \\
         \hline
    \end{tabular}
    \caption{Results from S-PPR for California (CA), New York (NY), Texas (TX), and Florida (FL), the four largest states in the US. We display the top 5 URLs per state according to S-PPR scores.
    We can see there are common URLs across states, such as the CVS and Walgreens COVID-19 vaccine pages, but also state-specific programs in each set.}
    \label{tab:sppr-results}
\end{table}
In this step, we select the URL candidates that external annotators will manually label for vaccine intent.
First, we filter out URLs that do not begin with ``http'', which leaves out URLs that are ads, internal links to other Microsoft verticals (e.g., News, Videos), and telephone numbers.
Then, we order the remaining URLs in each state according to their S-PPR scores, from highest to lowest.
We keep the \textit{union} over states of their top 100 URLs as our set of URL candidates, resulting in 2,483 candidates.
The number of URLs we have from taking the union over states is much lower than the number of states multiplied by 100, since there is overlap between states.
For example, the COVID-19 vaccine page for Walgreens is one of the most common URLs, appearing in the top 100 for all 50 states with an average ranking of 5.46 (where 0 indicates top-ranked).
However, there is also substantial heterogeneity in top URLs across states, reflecting state-specific vaccine programs and policies (Table \ref{tab:sppr-results}).
By constructing separate graphs and running S-PPR per state, our approach is uniquely able to capture this state-specific heterogeneity.
In Section \ref{sec:methods-bias-classifier}, we explore how alternative approaches that use a combined graph for multiple states severely hurt performance for small states.

Before presenting our URL candidates to annotators, we perform additional post-processing on the candidates in the union set.
First, we identify highly similar URL patterns that appear, such as the CVS store locator for the COVID-19 vaccine,\footnote{The CVS store locator for the COVID-19 vaccine is nearly identical across locations, such as \url{https://www.cvs.com/store-locator/cvs-pharmacy-locations/covid-vaccine/Hawaii/Honolulu} and \url{https://www.cvs.com/store-locator/cvs-pharmacy-locations/covid-vaccine/Montana/Billings}.} and only keep up to 5 URLs per pattern, so that our annotations are not overly repetitive.
This process reduces our set to 2,222 URLs.
Second, since we ask annotators to label URLs based on what they see when they click on the URL, we filter out URLs that now redirect to a different page.
For each URL candidate, we compute the normalized edit distance between the requested URL and the URL it redirects to by taking the Levenshtein distance divided by the length of the requested URL.
We keep URLs with a normalized edit distance less than or equal to 0.2, which keeps around 80\% of URLs.

\subsection{Annotation on Amazon Mechanical Turk} \label{sec:methods-amt}
\begin{figure}
    \centering
    \includegraphics[width=\textwidth]{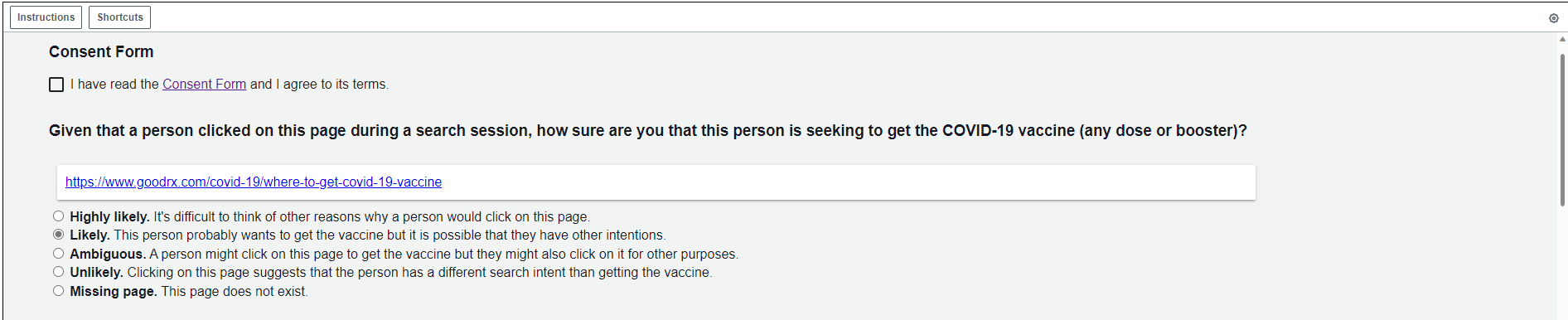}
    \caption{First question for Amazon Mechanical Turk task.}
    \label{fig:amt-url-q1}
\end{figure}
\begin{figure}
    \centering
    \includegraphics[width=\textwidth]{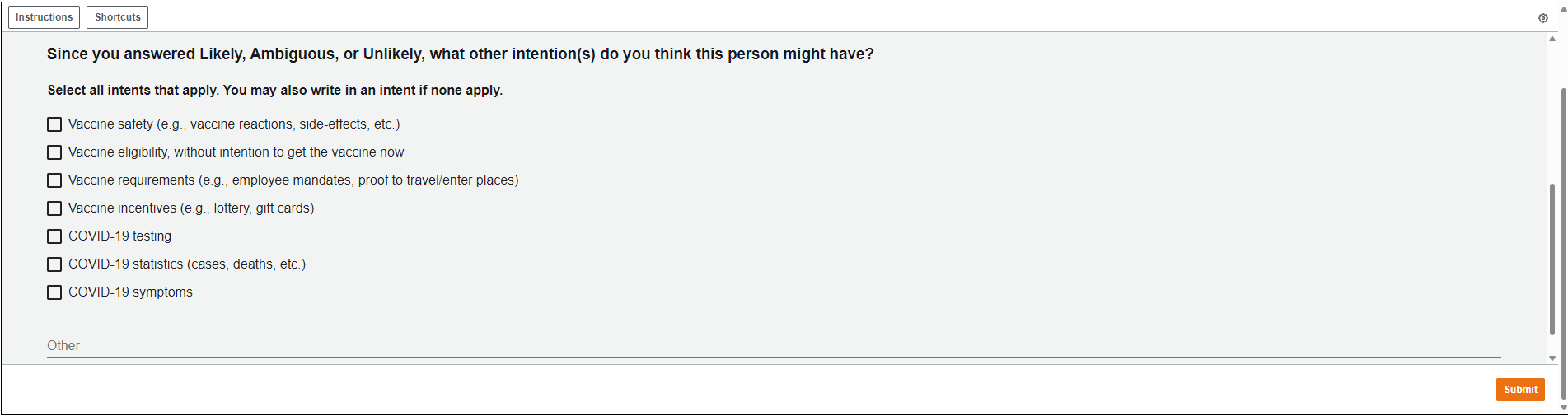}
    \caption{Second question for Amazon Mechanical Turk task.}
    \label{fig:amt-url-q2}
\end{figure}
\paragraph{Gathering annotations.} 
In this step, we present our URL candidates to annotators on Amazon Mechanical Turk (AMT). 
Our task instructs them to click on the presented URL and to answer based on what they see.
The first question we ask is, ``Given that a person clicked on this page during a search session, how sure are you that this person is seeking to get the COVID-19 vaccine (any dose or booster)?'' (Figure \ref{fig:amt-url-q1}).
We provide options from Highly Likely to Unlikely, as well as Missing Page.
If the annotator selected Likely, Ambiguous, or Unlikely, we ask them to indicate what other intention(s) the person might have, such as seeking information about vaccine safety or COVID-19 testing (Figure \ref{fig:amt-url-q2}). 
While we use the answers to the first question to construct our vaccine intent labels, we include the second question to encourage annotators to think broadly about vaccine-related searches, so that they would only label positively for vaccine intent if the URL seemed unambiguous.
To validate our vaccine intent queries, we also present a sample of queries to annotators.
To capture a diverse sample, we use the union over the top 5 and bottom 5 vaccine intent queries per state, after filtering out queries that were issued by fewer than 50 users (for privacy reasons) and sorting the remaining ones by their S-PPR scores.
This results in 227 vaccine intent queries to label.
In the query version of our task, we ask very similar questions to those shown for URLs, but replace language about clicking on the URL with issuing the query.

To make sure our questions were clear, we conducted two internal user studies, first with the authors doing a small pilot run, then with recruited colleagues at Microsoft doing a larger pilot run.
Our pilot studies allowed us to test the design of the questions, but our final vaccine intent labels were entirely based on the AMT labels that we received.
Our pilots also allowed us to estimate that each task (labeling a single URL or query) would take around 30 seconds. 
We set the compensation on AMT to \$0.15 per task, which corresponds to an hourly rate of around \$18.
Our AMT task was also approved by the Microsoft IRB Office, in a separate application from our approved analysis of Bing search logs, and we included a consent form in our instructions that annotators were required to read and sign before starting the task.

\paragraph{Annotation results.}
\begin{figure}
    \centering
    \includegraphics[width=7cm]{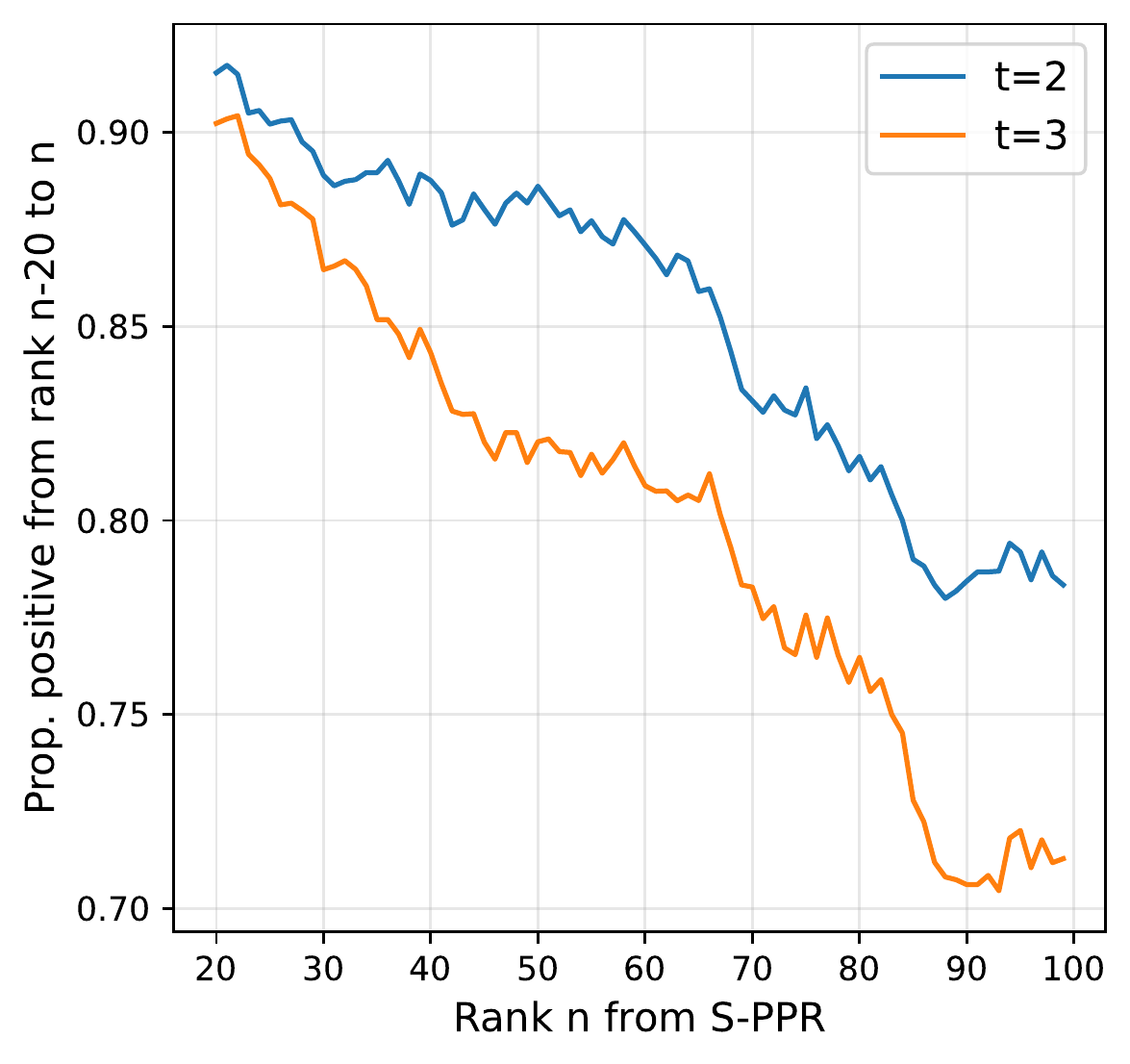}
    \caption{Comparison of S-PPR rank to the proportion of URLs around that rank that qualified for a positive vaccine intent label. We observe that these URLs (the top 100 from S-PPR) have a high positive rate overall: 86\% if we require 3 positive annotations ($t=3$), 90\% if we require 2 ($t=2$). Furthermore, the worse the S-PPR rank, the lower the positive rate, showing that our S-PPR technique is well-calibrated to predict vaccine intent.}
    \label{fig:amt-vs-sppr}
\end{figure}
For each URL, we first present it to three annotators. 
If all three give it a positive label (i.e., Highly Likely or Likely), then we label this URL as vaccine intent.
If two give it a positive label and one does not, we consider this a ``non-consensus'' URL, and we assign it to one more annotator.
If that annotator gives it a positive label, then we also label this URL as vaccine intent.
In other words, we require vaccine intent URLs to receive three positive annotations.
With this relatively strict bar, we still find that a large majority (86\%) of our URL candidates are labeled as vaccine intent.
Furthermore, we observe a clear relationship between S-PPR rank and the percentage labeled as vaccine intent: for example, around 90\% of URLs from ranks 0 to 20, around 81\% of URLs from ranks 40-60, and around 71\% of URLs from ranks 80 to 100 (Figure \ref{fig:amt-vs-sppr}).
We also find a very high positive rate among the vaccine intent queries that we tested.
Using the same annotation process and requirement of three positive labels, we find that 96\% of the vaccine intent queries we test are labeled as true vaccine intent.
The ones that are not seem to be mislabeled due to noise and our high bar for inclusion, since on inspection, they do seem to unambiguously communicate vaccine intent (e.g., [covid vaccines walgreens]).\footnote{Out of the 227 vaccine intent queries we evaluated, 9 did not receive three positive annotations: [covid vaccines walgreens], [cvs covid vaccine-19], [meijer covid vaccine], [meijers covid vaccine], [publix.com/covidvaccine], [walgreens.com covid vaccines], [walmart covid shots], [walmart covid vaccine walk-in], [walmart vaccine/covid].}

For additional quality control, we review the URLs labeled as vaccine intent and find that the majority of them seem correct. 
We remove a small number of URLs that still seem slightly ambiguous, e.g., the COVID-19 page for a jurisdiction that may not be vaccine-specific. 
In our review, we also find common URL patterns, such as COVID-19 vaccine store locators for CVS or Walmart.
We define regular expressions matching these patterns so that we can detect these types of URLs too, even if they are not in our predefined vaccine intent set.

\subsection{Graph neural networks for expansion} \label{sec:methods-gnn}
Since manual annotation is expensive, we wish to augment our manual efforts by training machine learning models on the AMT labels, then use the models to expand our set of vaccine intent URLs.
We formulate this problem as semi-supervised node classification on a graph, since the URLs are nodes in the query-click graph and we are trying to predict whether a URL indicates vaccine intent or not, given labels for a subset of URLs.
To solve this problem, we use graph neural networks\cite{kipf2017gcn} (GNNs), which are a powerful class of machine learning models for graph-structured data that naturally incorporate graph structure and node features into prediction.

\paragraph{Training GNNs to predict vaccine intent.}
Our model consists of two character-level convolutions (CNN), followed by three graph convolutions (GCN), followed by a final linear layer and sigmoid activation that produces the model prediction as a probability between 0 and 1.
We use the character-level CNN to capture textual information in the queries and URLs, since text can be informative for this problem (e.g., the appearance of ``vaccine'' or ``vax'').
Character-level representations are more natural for URLs, which often join words through hyphens or concatenation, or use abbreviations or truncated words.
Character-level representations also allow us to account for typos in queries.
The graph convolutions then allow us to learn representations of URLs that draw from the representations of their neighboring queries, which draw from the representations of their neighboring URLs, and so on.
In this way, we can capture ``similar'' URLs in embedding space (similar in terms of both text and graph structure), which allows us to learn embeddings that are predictive of vaccine intent.

Given the query-click graph for a state, we label a URL node as 1 if the URL was labeled as vaccine intent from AMT (following the inclusion criteria described in the previous section) or if it matches one of the regular expressions we identified for vaccine intent URLs.
We label it as 0 if at least two AMT annotators gave the URL negative labels (``Ambiguous'' or ``Unlikely'') or if it matches a regular expression we identified as \textit{not} vaccine intent, such as general store locators for pharmacies. 
To train and test our model, we randomly split the URL labels into a train set (60\%), validation set (15\%), and test set (25\%). 
We train the model on the train set, iteratively updating model parameters with gradient descent on the train loss (cross-entropy loss) and evaluating its loss on the validation set. We continue training until the model's validation loss is no longer improving.
Finally, we evaluate the model's performance on the held-out test set using area under the receiver operating characteristic curve (AUC), a standard metric in machine learning.
In Section \ref{sec:methods-bias-classifier}, we additionally evaluate the model's true and false positive rates, which are central to evaluating the model's bias (Section \ref{sec:methods-bias-decomp}).

However, some states have much smaller graphs, and therefore, fewer positive and negative labels.
For example, for the state of Wyoming, we only have 245 positive and 276 negative URLs.
We find that with such few labels, the model cannot adequately learn how to predict vaccine intent, with AUCs far below those of large states (Table \ref{tab:gnn-pretrain}).
Additionally, as we show below, we find that joining state graphs into one combined graphs also results in worse performance for smaller states, since larger states' labels and query-click patterns dominate.
Instead, our key insight is that we can retain state-level graphs and models, but we \textit{pre-train} the model for smaller states on S-PPR rankings (Section \ref{sec:methods-sppr}), which we have for many more URLs than we have labels for.
Our intuition is that S-PPR already performed remarkably well at predicting vaccine intent, as we showed in our annotation results (Section \ref{sec:methods-amt}).
Furthermore, S-PPR rankings do not require any additional manual labels; we derive them entirely from our initial vaccine intent queries, which were automatically labeled using regular expressions.
In practice, before training the model on the URL labels from AMT and regular expressions, we train the model to predict the URLs' S-PPR rankings that we derived in Step 1.
Since S-PPR rankings become less meaningful in the long tail of URLs, we focus on the top $K = \max(1000, q_{\max})$ S-PPR rankings, where $q_{\max}$ is the maximum rank (where lower rank corresponds to higher S-PPR score) of the last seed set query.
This pre-training encourages the model to learn URL representations that are predictive of S-PPR rankings, which we find help substantially with the ultimate task of predicting vaccine intent.

\paragraph{Evaluating GNN performance.}
We evaluate model performance on the held-out test set by computing its AUC, which captures how well the model trades off between its true positive rate and false positive rate.
Furthermore, to account for randomness from model training and data splitting, we run 10 random trials for every model/state, where in each trial, we re-split the URL labels into train, validation, and test sets, retrain the model on the train set (stopping based on the validation loss), and re-evaluate the model's final performance on the test set.

\begin{table}
    \centering
    \begin{tabular}{c|c|c|c}
        State & Number of nodes & AUC w/o pre-train & AUC w/ pre-train \\
        \hline 
        Wyoming & 752865 & 0.741 (0.146) & 0.951 (0.014) \\
        Alaska & 909357 & 0.796 (0.187) & 0.921 (0.074) \\
        Delaware & 1269327 & 0.864 (0.134) & 0.968 (0.007) \\
        Montana & 1533071 & 0.857 (0.139) & 0.978 (0.011) \\
        Connecticut & 4407722 & 0.987 (0.005) & 0.984 (0.008) \\
        Tennessee & 7712443 & 0.991 (0.003) & 0.990 (0.003)
    \end{tabular}
    \caption{Effects of pre-training on S-PPR rankings for 6 selected states that vary in size. We report the mean and standard deviation of AUC on the held-out test set over 10 random trials.}
    \label{tab:gnn-pretrain}
\end{table}
\begin{figure}
    \centering
    \includegraphics[width=10cm]{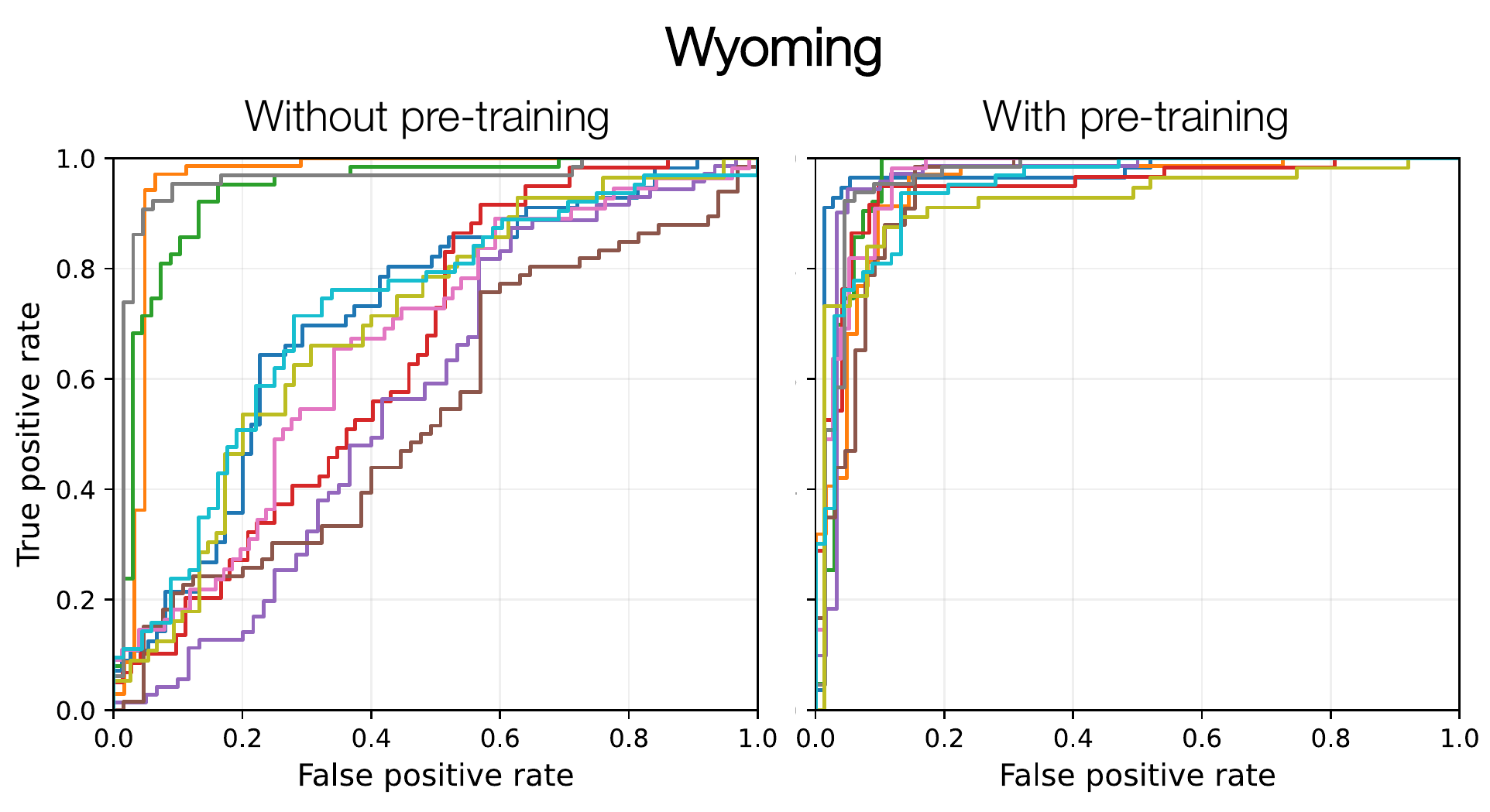}
    \caption{Visualizing the ROC curve for Wyoming, one of the smallest states. Each line represents a random trial. Without pre-training on S-PPR rankings, performance is unstable across trials, with some trials performing poorly (left). Pre-training stabilizes performance (right) and significantly improves AUC for smaller states (Table \ref{tab:gnn-pretrain}).}
    \label{fig:wyoming_aucs}
\end{figure}
First, we select six representative states, chosen to vary in graph size and US region, to test the effect of pre-training on S-PPR rankings.
We find that pre-training significantly improves performance for the smaller states; for example, the mean AUC for Wyoming increases from 0.74 to 0.95 (Table \ref{tab:gnn-pretrain}, Figure \ref{fig:wyoming_aucs}).
Specifically, due to the low number of URL labels for smaller states, we observe great variance in the model's performance if we do not pre-train the model, leading to some trials that perform well and some that perform poorly.
Performance becomes far more stable for smaller states after we incorporate the pre-training objective.
We find that pre-training seems unnecessary for the larger states, such as Connecticut and Tennesssee, where we are already achieving high AUCs above 0.98.
So, we set a generous cutoff of 5,000,000 nodes (still larger than the graph size for Connecticut) and we pre-train all states with fewer than 5,000,000 nodes in our data, of which there are 26.
After incorporating pre-training for these smaller states, we are able to achieve AUCs above 0.90 for all 50 states and above 0.95 for 45 states (Figure \ref{fig:results-across-states}). 

\begin{figure}
    \centering
    \includegraphics[width=\linewidth]{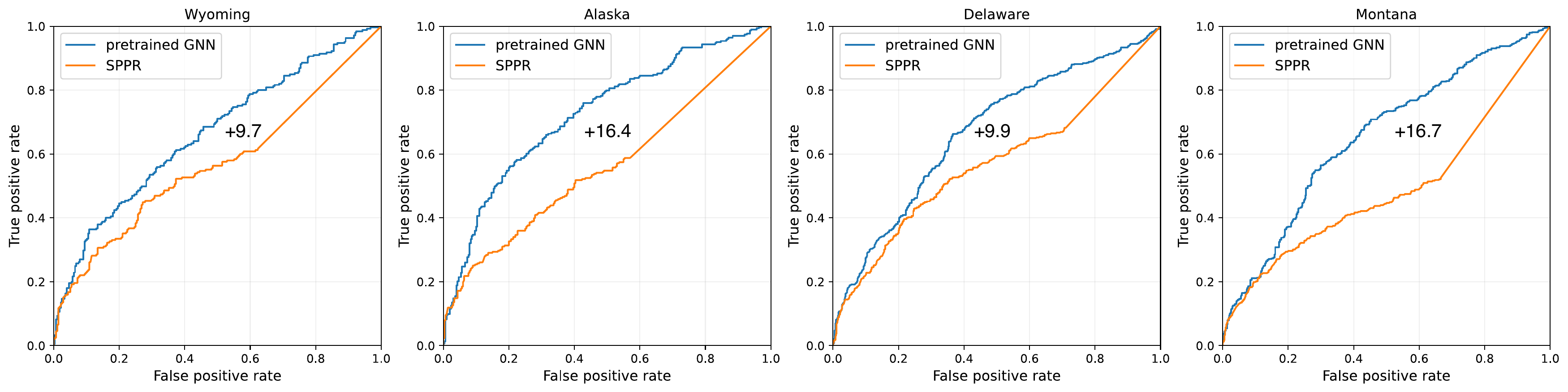}
    \caption{Comparing S-PPR vs. GNN pre-trained on S-PPR rankings (GNN-PPR). Despite only being trained on S-PPR rankings, GNN-PPR consistently outperforms S-PPR at predicting URL labels from AMT (unseen by both). In the figures, the number between the curves indicates the increase in AUC from S-PPR to GNN-PPR.}
    \label{fig:sppr-vs-gnn-ppr}
\end{figure}
As a supplementary analysis, we can also use AUC to evaluate the predictive performance of S-PPR alone and GNN-PPR, i.e., the GNN pre-trained on S-PPR rankings \textit{before} it is also trained on AMT labels. 
Here, we evaluate on \textit{all} AMT labels, since none of them were used in constructing S-PPR or GNN-PPR scores.
In fact, evaluating on AMT labels is particularly challenging, since we chose to label only the top-ranked URLs according to S-PPR, so we are asking S-PPR to distinguish between URLs that it already considers similar.
We conduct this experiment on the 26 smaller states for which we pre-trained our GNNs.
First, we find across these states that S-PPR still performs better than random, with a mean AUC of 0.569, which complements our annotation results showing that even within its top-ranked URLs, S-PPR rankings still correlate with true rates of vaccine intent labels (Figure \ref{fig:amt-vs-sppr}).
Second, we find that GNN-PPR consistently \textit{outperforms} S-PPR by 10-15 points, with a mean AUC of 0.675 (Figure \ref{fig:sppr-vs-gnn-ppr}).
This is somewhat surprising, since GNN-PPR was only trained to predict S-PPR rankings, without any additional labels.
We hypothesize that GNN-PPR outperforms S-PPR because, unlike S-PPR, the GNN can incorporate textual information from URLs and queries, in addition to graph structure.
So, while S-PPR incorrectly upweights high-traffic URLs such as facebook.com that are often reached on random walks starting from the vaccine intent queries, GNN-PPR recognizes that these URLs do not look like the rest of high-ranking URLs and correctly excludes them.
However, in order to achieve this difference between S-PPR and GNN-PPR, it is important not to overfit on S-PPR. 
So, we employ early stopping during pre-training; that is, we train the GNN on S-PPR rankings until they achieve a correlation of 0.8 and then we stop pre-training.

Our evaluation results demonstrate that our GNNs are able to accurately predict vaccine intent labels in all 50 states, which is essential as we use our GNNs to discover new vaccine intent URLs.
Furthermore, the supplemental analysis shows that due to the expressive power of the GNN (with character-level CNN) and the predictive power of S-PPR from a well-designed seed set, we can achieve decent performance without \textit{any} labels at all.
These methods, which should be explored more deeply in future work, may be useful in a zero-shot context, allowing lightweight, effective prediction before acquiring any labels.

\paragraph{Discovering new vaccine intent URLs.}
Finally, we use our trained GNNs to identify new vaccine intent URLs.
We apply our GNNs to predict scores for all unlabeled URLs within the top $K$ URLs according to S-PPR ranking (again, with $K = \max(1000, q_{\max})$).
However, in order to decide which new URLs to include as vaccine intent, we need to determine a score threshold.
Our goal is to set the threshold such that any URL that scores above it is very likely to truly be vaccine intent (i.e., we want to have high precision).
Borrowing the idea of ``spies'' from positive-unlabeled learning,\cite{bekker2020pu} our idea is to use the held-out positive URLs in the test set to determine where to set the threshold.
We consider two thresholds: (1) $t_{\text{med}}$, the median score of the held-out positive URLs, and (2) $t_{\text{prec}}$, the minimum threshold required to achieve precision of at least 0.9 on the held-out test set.
Then, we only include URLs that pass both thresholds in at least 6 out of the 10 random trials (where, as described before, we reshuffle the data and retrain the model per trial).

Our method is similar to common ``big data'' approaches that, due to the scale of unlabeled data, seek to manually annotate a subset of data, train machine learning models to accurately predict those labels, then use those models to label the rest of the data.\cite{grimmer2021social,card2022pnas,lundberg2022social,franchi2023police}
We extend this approach with special attention to the classification threshold, setting it high so that we can ensure high precision among the new URLs that we discover.
Even with this high threshold, we discover around 11,400 new URLs, increasing our number of vaccine intent URLs by 10x.
\begin{table}[]
    \small
    \centering
    \begin{tabular}{p{14cm} | p{1cm} | p{1cm}}
    URL & $t_{\text{med}}$ & $t_{\text{prec}}$ \\
    \hline 
    \url{https://www.chesco.org/4836/61876/COVID-Authorized-Vax} & 7 & 10 \\[0.1cm]
    \url{https://patch.com/new-jersey/princeton/all-information-princeton-area-covid-vaccine-sites} & 9 & 10 \\[0.6cm]
    \url{https://dph.georgia.gov/locations/spalding-county-health-department-covid-vaccine} & 9 & 10 \\[0.6cm]
    \url{https://www.abc12.com/2021/04/22/whitmer-says-covid-19-vaccine-clinics-like-flint-church-are-key-to-meeting-goals/} & 7 & 10 \\[0.6cm]
    \url{https://www.delta.edu/coronavirus/covid-vaccine.html} & 10 & 10 \\[0.1cm]
    \url{https://www.lewistownsentinel.com/news/local-news/2021/01/scheduling-a-virus-vaccine-appointment/} & 9 & 10 \\[0.6cm]
    \url{https://www.laconiadailysun.com/news/local/covid-vaccine-clinics-at-lrgh-franklin-now-open-to-public/article_aa4b67e0-601a-11eb-a889-1bd4e6c83de1.html} & 6 & 10 \\[1.1cm]
    \url{https://www.insidenova.com/headlines/inside-woodbridges-new-mass-covid-19-vaccination-site-the-lines-keep-moving/article_eca45b88-8db0-11eb-a649-4bbeccd82cc3.html} & 9 & 10 \\[1.1cm]
    \url{https://www.keloland.com/news/healthbeat/coronavirus/avera-opens-covid-19-vaccine-clinic/} & 10 & 9 \\[0.6cm]
    \url{https://bangordailynews.com/2021/04/06/news/maine-to-kick-off-statewide-mobile-covid-19-vaccine-clinics-in-oxford-next-week-sk6sr8zcdk/} & 8 & 9 \\[0.6cm]
    \url{https://morgancounty.in.gov/covid-19-vaccinations/} & 9 & 10 \\[0.1cm]
    \url{https://www.firsthealth.org/specialties/more-services/covid-19-vaccine} & 10 & 10 \\[0.1cm]
    \url{https://healthonecares.com/covid-19/physician-practices/covid-19-vaccine-information.dot} & 9 & 10 \\[0.6cm]
    \url{https://patch.com/florida/stpete/drive-thru-covid-19-vaccine-sites-open-florida} & 9 & 10 \\[0.6cm]
    \url{https://vaccinate.iowa.gov/eligibility/} & 7 & 10 \\[0.1cm]
    \url{https://www.baynews9.com/fl/tampa/news/2021/03/17/new-walk-in-vaccine-site-at-tpepin-hospitality-centre-opens-today} & 10 & 10 \\[0.6cm]
    \url{https://www.doh.wa.gov/Emergencies/COVID19/VaccineInformation/FrequentlyAskedQuestions} & 10 & 10 \\[0.6cm]
    \url{https://www.emissourian.com/covid19/vaccine-registration-open-for-franklin-county/article_3638f7a0-5769-11eb-9bba-3f2611173784.html} & 10 & 10 \\[0.6cm]
    \url{https://www.fema.gov/press-release/20210223/maryland-open-covid-19-vaccination-center-waldorf-fema-support} & 10 & 10 \\[0.6cm]
    \url{https://kingcounty.gov/depts/health/covid-19/vaccine/forms.aspx} & 10 & 10 \\
    \end{tabular}
    \caption{A random sample (\texttt{random\_state=0}) of 20 URLs from GNN. $t_{\text{med}}$ and $t_{\text{prec}}$ indicate how often the URL passed the median cutoff and precision cutoff, respectively, out of the 10 trials.}
    \label{tab:gnn-urls}
\end{table}
In Table \ref{tab:gnn-urls}, we provide a uniform random sample of the URLs that our GNNs discovered.
The majority of them seem to express vaccine intent, with several news stories about new vaccine clinics and information about vaccine appointments.
In the following section, we also evaluate the impact of adding these URLs discovered by GNNs on our ability to estimate regional vaccine intent rates.
We find that the new URLs not only increase our coverage of vaccine intent users by 1.5x but also further improve our agreement with reported vaccination rates from the CDC (Table \ref{tab:pipeline-results}).

\section{Estimating vaccine intent rates \& correcting for bias} \label{sec:methods-bias}
In this section, we discuss how we use our classifier to estimate regional rates of vaccine intent and how we correct for and evaluate sources of bias in our estimates.

\subsection{Decomposition of bias} \label{sec:methods-bias-decomp}
For a given individual, let $v \in \{0, 1\}$ indicate whether they actually had vaccine intent (up to a certain time) and $\hat{v} \in \{0, 1\}$ indicate whether our classifier labels them as having vaccine intent.
Furthermore, let $r$ represent the individual's home region, such as their state or county.
We would like to estimate the regional vaccine intent rate, $\Pr(v | r)$, but we do not have access to $v$, only to $\hat{v}$.
To understand how simply using $\hat{v}$ in place of $v$ may bias our estimates, let us relate $\Pr(\hat{v} | r)$ to $\Pr(v | r)$.
First, we introduce another variable $b$, which represents whether the individual is a Bing user.
Note that $\hat{v} = 1$ implies that $b = 1$, since our classifier can only identify vaccine intent from users who appear in Bing search logs.
With these variables, we have
\begin{align}
    \Pr(\hat{v}=1 | r) = \underbrace{\Pr(b=1|r)}_{\text{Bing coverage of $r$}}[&\Pr(v = 1 | r) \underbrace{\Pr(\hat{v}=1 | b=1, v=1, r)}_{\text{Classifier TPR for $r$}}\\+ &\Pr(v = 0 | r) \underbrace{\Pr(\hat{v}=1 | b=1, v=0, r)}_{\text{Classifier FPR for $r$}}], \nonumber
\end{align}
where TPR and FPR are the true and false positive rates, respectively.
$\Pr(b=1|r)$ represents the probability that an individual from region $r$ is a Bing user, i.e., the Bing coverage of $r$.
Incorporating $b$, $v$, and $r$ into $\Pr(\hat{v}|b, v, r)$ reflects all of the factors that affect whether the classifier predicts vaccine intent.
As discussed, if the user is not a Bing user ($b=0$), then the probability is 0, so we only consider the $b=1$ case.
If $v=1$, predicting $\hat{v}=1$ would be a true positive; if $v=0$, it would be a false positive.
Conditioning $\hat{v}$ on region $r$ reflects the possibility that individuals from different regions may express vaccine intent differently and the classifier may be more prone to true or false positives for different regions.
Finally, we make the assumption here that $b \perp v | r$; that is, conditioned on the individual's region, being a Bing user and having vaccine intent are independent.
This misses potential within-region heterogeneity, but to mitigate this in practice, we focus on fine-grained regions (ZIP code tabulation areas, Section \ref{sec:methods-bing-coverage}).

Based on this decomposition, we can see that if Bing coverage, TPR, and FPR are uniform across regions, then $\Pr(\hat{v}|r)$ will simply be a linear function of $\Pr(v|r)$.
Unfortunately, we know that Bing coverage is not uniform.
However, we observe $b=1$ and can assign users to regions, so we can estimate Bing coverage per region and correct by inverse coverage.
Thus, our estimate corresponds to a coverage-corrected predicted vaccine intent rate, $\vi{r} = \frac{\Pr(\hat{v}=1|r)}{\Pr(b=1|r)}$.
If we refer to the true vaccine intent rate as $p(v, r)$, then we can see that $\vi{r}$ is a linear function of $p(v, r)$ when TPR and FPR are uniform:
\begin{align}
    \frac{\Pr(\hat{v}=1|r)}{\Pr(b=1|r)} &= \Pr(v=1|r)\text{TPR} + (1-\Pr(v=1|r))\text{FPR} \label{eqn:tpr-fpr} \\ 
    \vi{r} &= \text{FPR} + (\text{TPR}-\text{FPR})p(v, r). 
    \nonumber
\end{align}
Furthermore, if FPR is low, then $\vi{r}$ is approximately proportional to $p(v, r)$.
Thus, our first two strategies for addressing bias in our estimates are:
\begin{enumerate}
    \item Estimate Bing coverage per region and weight by inverse coverage (Section \ref{sec:methods-bing-coverage}),
    \item Evaluate whether we observe similar TPRs and FPRs across regions and whether FPRs are close to 0 (Section \ref{sec:methods-bias-classifier}).
\end{enumerate}
These efforts are our first two lines of defense against bias.
After this, we can furthermore compare our final vaccine intent estimates to established  data sources, such as the CDC's reported vaccination rates (Section \ref{sec:methods-cdc}) and Google search trends (Section \ref{sec:methods-google}).

\subsection{Coverage-corrected vaccine intent rates} \label{sec:methods-bing-coverage}
\paragraph{Estimating Bing coverage.}
Our goal here is to estimate $\Pr(b = 1|r)$, the probability that an individual from region $r$ is a Bing user.
We focus on ZIP Code Tabulation Areas (ZCTAs) as our fine-grained notion of regions; for example, there are approximately 10x more ZCTAs in the US than counties.
ZIP codes are the most granular geographic area that we can assign Bing users to, since we have, for most Bing queries, a record of which ZIP code the query came from.
We focus on ZCTAs, which are ``generalized areal representations'' of ZIP codes, since they are a unit that the Census tracks and provides demographic information about.\cite{census_zctas}

We consider a Bing user ``active'' in a given month if they issue at least 30 queries in that month. For
most (over 90\%) of queries, Bing estimates the ZIP code, county, and state from which the query originates.
Based on an active user's query-level ZIP codes from the month, we assign the user to their mode ZIP code
if the mode accounts for at least 10 and at least 25\% of these queries (with the same rules for assigning county and state).
We assume the mode is the user’s likeliest home location from this month and include these additional requirements to avoid assigning users to locations that they just happened to visit and query from, but do not live in. 
Focusing on active users with a larger number of queries also improves our ability to reliably assign a user to a location.
We estimate $N(b, z)$, the number of active Bing users from ZCTA $z$, as the average number of active users assigned to $z$ over the months in our study period (February to August 2021).\footnote{In most instances, there is a one-to-one mapping from ZCTA to ZIP code. For the ZCTAs that contain multiple ZIP codes, we set $N(b, z)$ to the sum of average user counts over those ZIP codes.}
We also acquire $N(z)$, the population size of $z$, from the 2020 5-year American Community Survey.\cite{census_acs}
Finally, we estimate the ZCTA's coverage $\Pr(b=1|z)$ as $\frac{N(b,z)}{N(z)}$.

\paragraph{Computing vaccine intent rates with inverse coverage.}
Recall that our goal is to estimate $\vi{z} = \frac{\Pr(\hat{v}=1|z)}{\Pr(b=1|z)}$.
To estimate $\Pr(\hat{v}=1|z)$, we apply our vaccine intent classifier to all queries and clicks of active Bing users.
This produces $N(\hat{v}, z)$, the number of active Bing users from $z$ for whom we detect vaccine intent. Then,
\begin{align}
    \vi{z} = \frac{\Pr(\hat{v}=1|z)}{\Pr(b=1|z)} = \frac{\frac{N(\hat{v},z)}{N(z)}}{\frac{N(b,z)}{N(z)}} = \frac{N(\hat{v}, z)}{N(b, z)}. \label{eqn:vi-z}
\end{align}
This is an intuitive result: our estimate for the vaccine intent rate in $z$ is the number of active Bing users with predicted vaccine intent divided by the total number of active Bing users. 
We can use these ZCTA-level rates to characterize demographic trends, such as by computing correlations between $\vi{z}$ and some demographic of $z$, such as its percentage aged 65 and over.
It is also often useful to aggregate $\vi{z}$ over sets of ZCTAs, e.g., to the state or county-level.
To compute the vaccine intent rate for a set $Z$ of ZCTAs, we simply take the population-weighted average:
\begin{align}
    \vi{Z} = \frac{\sum_{z \in Z} N(z) * \vi{z}}{\sum_{z \in Z} N(z)}. \label{eqn:vi-set}
\end{align}
For example, we use estimated state and county vaccine intent rates to compare against reported vaccination rates from the CDC (Section \ref{sec:methods-cdc}).
This population-weighted average is equivalent to post-stratification,\cite{little1993poststrat} a common technique for adjusting non-representative survey responses to match known population totals, where we treat each ZCTA as a post-stratum.

\subsection{Bias in vaccine intent classifier} \label{sec:methods-bias-classifier}
Our primary source of bias is uneven Bing coverage, which we found can vary by more than 2x across ZCTAs. 
However, after correcting for Bing coverage, we also want to know that our classifier does not significantly contribute to additional bias. 
To do this, we must establish that our classifier’s true and false positive rates do not vary significantly or systematically across regions. 
The challenge is that we cannot perfectly evaluate our classifier’s true or false positive rates, because we do not know all true positives or true negatives. However, we can approximate these metrics based on the labeled URLs that we do have and furthermore make methodological decisions that encourage similar performance across groups.

\paragraph{Step 1: Personalized PageRank for URL candidates.}
Recall that in the first step of our pipeline, we generate URL candidates for annotation by propagating labels from vaccine intent queries to unlabeled URLs in query-click graphs (Section \ref{sec:methods-sppr}).
Since all URL candidates then go through manual inspection in Step 2, we do not have to worry about the false positive rate at this stage.
However, we do need to worry about the true positive rate (i.e., recall).
For example, if we only kept COVID-19 vaccine registration pages for pharmacies that are predominantly in certain regions, then we could be significantly likelier to detect true vaccine intent for certain states over others.
So, through the design and evaluation of our label propagation techniques, we aim to
ensure representativeness in vaccine intent across the US.

The most important design decision is that we construct query-click graphs \textit{per state}, then we run S-PPR per graph and take the union over states of top URLs as our set of URL candidates. 
Running this process separately for each state allows us to capture how vaccine intent varies regionally, with state-specific programs and websites for scheduling the vaccine (Table \ref{tab:sppr-results}).
To demonstrate the risks of not using a state-specific approach, we try an alternative approach where we construct a joint graph that combines the queries and clicks for 6 states, chosen to vary in graph size and US region (the same 6 states as those used in the pre-training experiments of Table \ref{tab:gnn-pretrain}).
To represent our union approach, we take the union over these 6 states of the top 200 URLs per state, which results in 935 URLs.
We compare this to a joint approach, where we take the top 935 URLs from running S-PPR on the joint graph.
To evaluate each approach, we compute the proportion of each state’s top $N$ URLs that are kept across different values of $N$.
While we cannot be sure that every URL in the state’s top $N$ is truly vaccine intent, from our annotation results, we saw high positive rates for top-ranking URLs (Figure \ref{fig:amt-vs-sppr}), so we would like to see similar recall at these ranks.

\begin{figure}
    \centering
    \includegraphics[width=11cm]{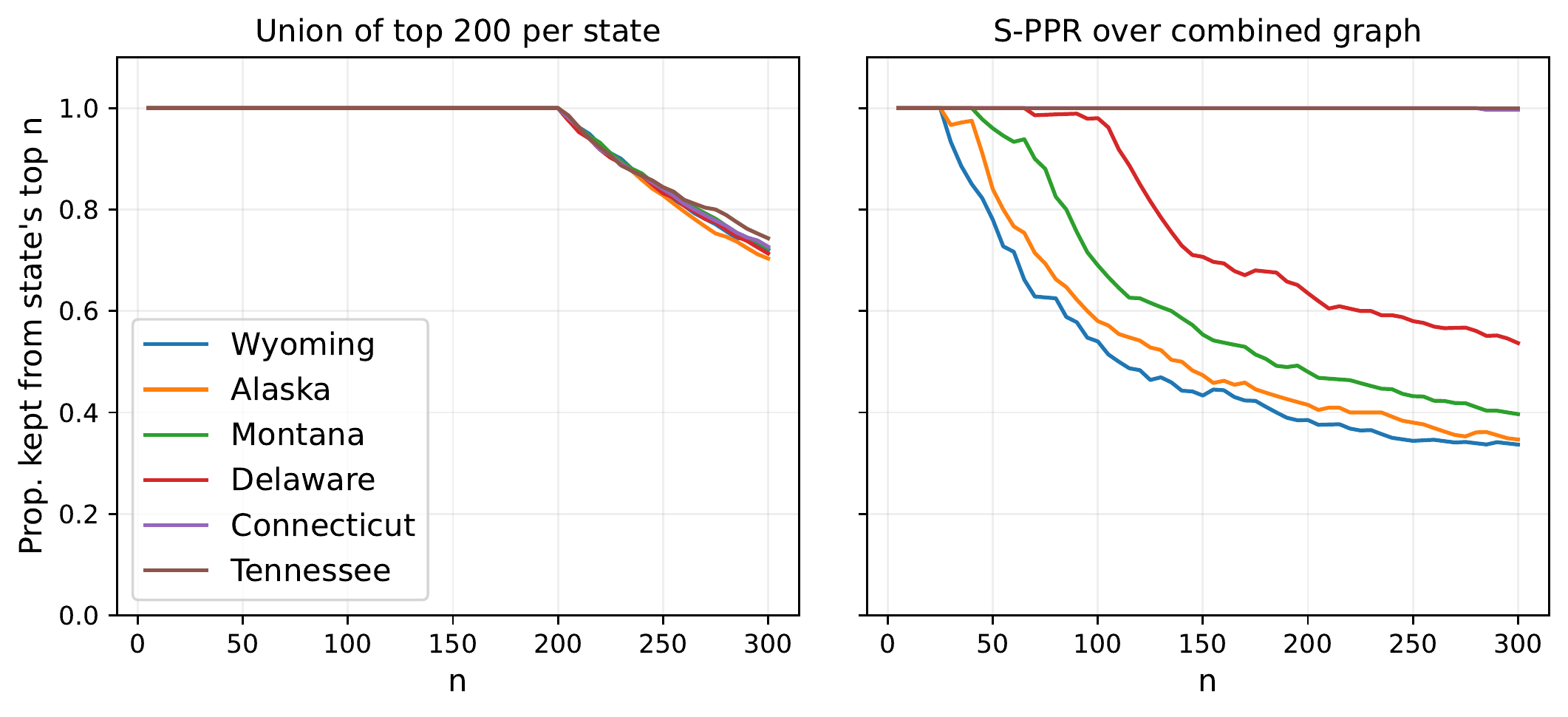}
    \caption{Comparing our union-over-states (left) to a combined graph approach (right) for generating URL candidates.}
    \label{fig:union-vs-combined}
\end{figure}
By design, our union-over-states approach ensures equivalent, 100\% recall up to $N = 200$ for all states (Figure \ref{fig:union-vs-combined}, left).
In comparison, we find that the joint approach yields different recalls as early as $N = 30$, with much higher recall for large states than small states (Figure \ref{fig:union-vs-combined}, right).
For example, it keeps less than 80\% of Wyoming’s URLs around rank 50 and less than 60\% around rank 100, while keeping 100\% of Tennessee’s throughout.
Furthermore, even past $N = 200$, where our union-over-states approach no longer has guarantees, we find that it still achieves far more similar recalls between states than the joint approach.
Thus, our design decisions enable similar recalls between states, which helps to reduce downstream model bias.
We also cast a wide net when constructing query-click graphs (taking all queries and clicks that co-occur in a session with any query that includes a COVID-19 or vaccine-related word), which may also improve recall and reduce bias, in case our choice of
initial keywords was not representative of all vaccine intent searches across the US.

\paragraph{Step 3: expanding vaccine intent URLs with GNNs.}
\begin{figure}
    \centering
    \includegraphics[width=\linewidth]{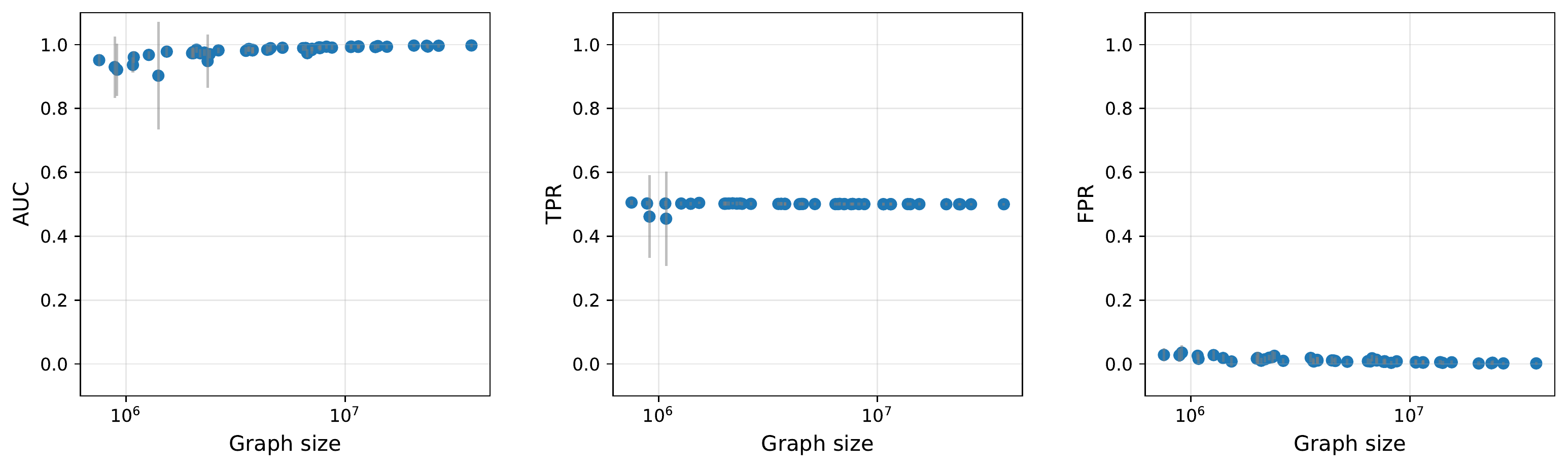}
    \caption{AUCs (left), true positive rates (middle), and false positive rates (right) across states. Metrics evaluate GNN performance on the held-out test set. Each dot represents a single state, with its y-coordinate representing the mean metric over 10 trials and grey bars indicating standard deviation.}
    \label{fig:results-across-states}
\end{figure}

In the third step of our pipeline, we use GNNs to expand our set of vaccine intent URLs beyond the manually labeled ones. 
We would like to see that the performance of GNNs is strong across states, to ensure that the GNN is not creating additional bias when expanding the URL set.
We showed in Section \ref{sec:methods-gnn} that, after incorporating pre-training on S-PPR rankings for smaller states, GNNs could achieve AUCs above 0.90 for all 50 states (Figure \ref{fig:results-across-states}, left).
The main metrics of interest when considering bias, however, are the true and false positive rates (TPRs and FPRs).
Unlike AUC, which is evaluated across decision thresholds, TPR and FPR depend on the chosen threshold $t$ above which data points are predicted to be positive.
In our setting, we set $t = \max(t_{\text{med}}, t_{\text{prec}})$, since we required new vaccine intent URLs to score above these two thresholds (in at least 6 out of 10 trials): (1) $t_{\text{med}}$, the median score of positive URLs in the test set and (2) $t_{\text{prec}}$, the minimum threshold required to achieve precision of at least 0.9 on the test set.
Then, we estimate TPR as the proportion of positive URLs in the test set that score above $t$ and FPR as the proportion of negative URLs in the test set that score above $t$.

We find that TPR is highly similar across states and hovers around 0.5 for all states (Figure \ref{fig:results-across-states}, middle).
This is because in almost all cases, $t_{\text{med}}$ is the higher of the two thresholds and thus the value of $t$, so the true positive rate lands around 0.5 since $t_{\text{med}}$ is the median score of the true positives.
FPR is also highly similar across states and very low (around 0.01; Figure \ref{fig:results-across-states}, right), which suggests that the quantity we estimate, $\vi{r}$, is not only a linear function of the true vaccine intent rate, $p(v, r)$, but also approximately proportional to it (Eq. \ref{eqn:tpr-fpr}).
The low FPR is encouraged but not guaranteed by our second threshold, $t_{\text{prec}}$. This threshold ensures that precision is over 0.9, which is equivalent to the false positive rate \textit{among the predicted positives} being below 0.1, which typically corresponds to low false positive rates over all true negatives (which is what FPR measures).
The GNN’s similar AUCs, TPRs, and FPRs across states, as well as the equivalent recalls in our label propagation stage, increase confidence that our classifier is not adding significant bias to our estimates.
In this section, we focused on states, since it was natural to evaluate performance per state due to the state-specific query-click graphs and models, and since we expect the expression of vaccine intent to vary most systematically per state due to state-specific vaccine programs and policies.
In the following section, we continue this analysis by comparing our final $\vi{r}$ estimates per state to CDC vaccination rates, but we also test out finer-grained evaluations, including vaccination rates over time and rates at the county-level.

\subsection{Comparison to CDC vaccination rates} \label{sec:methods-cdc}
\paragraph{Vaccination rates across states.}

The CDC releases vaccination rates at the state and county levels.
First, we compare against cumulative state-level vaccination rates. 
As our measure, we have $\vi{s}$ per state $s$, the cumulative vaccine intent rate in the state up to August 31, 2021, computed as described in Eq. \ref{eqn:vi-set}. 
On the CDC side, we use the cumulative proportion of population fully vaccinated\footnote{For cumulative metrics at the state and county-level, we compare to the CDC's percent fully vaccinated, which means
completing the second dose of a 2-dose series or completing the first dose of a single-dose series. We compare to percent fully vaccinated instead percent with at least one dose, since the CDC reported the latter can be overestimated: \url{https://www.cdc.gov/coronavirus/2019-ncov/vaccines/reporting-vaccinations.html}.} by August 31, 2021. 
We find a strong Pearson correlation between these cumulative rates, with $r = 0.86$ (Figure \ref{fig:classifier}c).
Notably, we find that the correlation drops to $r = 0.79$ if we do not correct for Bing coverage in our estimates and use a naive estimate instead that divides the total number of active users with vaccine intent by the total population size of the state (both summed over ZCTAs in the state), $\frac{\sum_z N(\hat{v}, z)}{\sum_z N(z)}$.
\begin{table}
    \centering
    \begin{tabular}{c|c|c}
         \textbf{Pipeline step} & \textbf{Correlation with CDC} & \textbf{Num vaccine intent users} \\
         \hline
        Only queries & 0.62 & 3.18M \\
        +manual URLs & 0.80 & 4.95M \\
        +manual and GNN URLs & 0.86 & 7.45M
    \end{tabular}
    \caption{Each step of our classification pipeline (Section \ref{sec:methods-classifier}) improves both our correlation with CDC state vaccination rates and our coverage of vaccine intent users.}
    \label{tab:pipeline-results}
\end{table}
We also find that each step of our classification pipeline improves the correlation with CDC (Table \ref{tab:pipeline-results}): if we only use the seed set queries identified by our regex, $r = 0.62$; if we use the queries plus the URLs identified from manual annotation, $r = 0.80$.
Furthermore, each step of our pipeline substantially increases the number of vaccine intent users we detect, which provides additional power for our downstream analyses. 

\paragraph{Vaccination rates over time.}
We also compare against vaccination rates over time. 
For a given state $s$ and day $t$, we can compute $\vi{s,t}$, the vaccine intent rate on day $t$; this is equivalent to Eq. \ref{eqn:vi-set}, except we only count users who showed their \textit{first} vaccine intent on day $t$ instead of the cumulative count of users who have shown vaccine intent up to day $t$.
We compare our metric to the daily proportion of individuals getting their first dose, since we expect the timing of the first vaccine intent to align more with the first dose than other doses, with a possible lag time.
We can calculate the daily proportion for first dose from the cumulative at-least-one-dose time series provided by the CDC (by subtracting the cumulative count for day $t$ from day $t-1$).
Furthermore, we apply one-week smoothing to both daily time series (taking the average from day $t-6$ to $t$, inclusive), to smooth out daily effects such as potential underreporting on the weekends.
For this analysis, we compare time series from April 1 to August 31, 2021, leaving out the months of February and March since we expect that vaccine intent was less correlated with actual vaccination rates early in the vaccine roll-out, since individuals would seek the vaccine but not be able to receive it yet (e.g., because they were not eligible or because there was not enough supply).
Then, we compute the Pearson correlation between the smoothed daily time series, allowing the CDC time series to lag behind the vaccine intent time series by $l = \{0, 1,\cdots, 21\}$ days.

For each state, we compute the maximum correlation possible using the optimal lag.
We observe strong temporal correlations, with correlations above $0.7$ for 48 states and a median correlation of $r = 0.89$.
We also observe a substantial lag between the vaccine intent and CDC time series, with a median optimal lag of 10 days (50\% CI, $7.5$-$14.5$); without any lag, the median correlation drops to $r = 0.78$.
In Figure \ref{fig:vi-cdc-over-time}, we visualize the vaccine intent and CDC vaccination time series for the 15 largest states in the US.
Almost all of the correlations are strong except for North Carolina, where the CDC time series shows an anomalous peak at the beginning of July 2021.\footnote{This peak seems to be an artifact of the CDC data, since it is so much larger and sharper than any other peak we see and cannot be easily explained by concurrent news or events.
These anomalies further motivate the need for complementary data sources, such as our classifier, that can also track vaccine seeking over time.}
We also see from this figure that the optimal lag varies across these states, but the CDC time series is always at least one week behind vaccine intent.
\begin{figure}
    \centering
    \includegraphics[width=\linewidth]{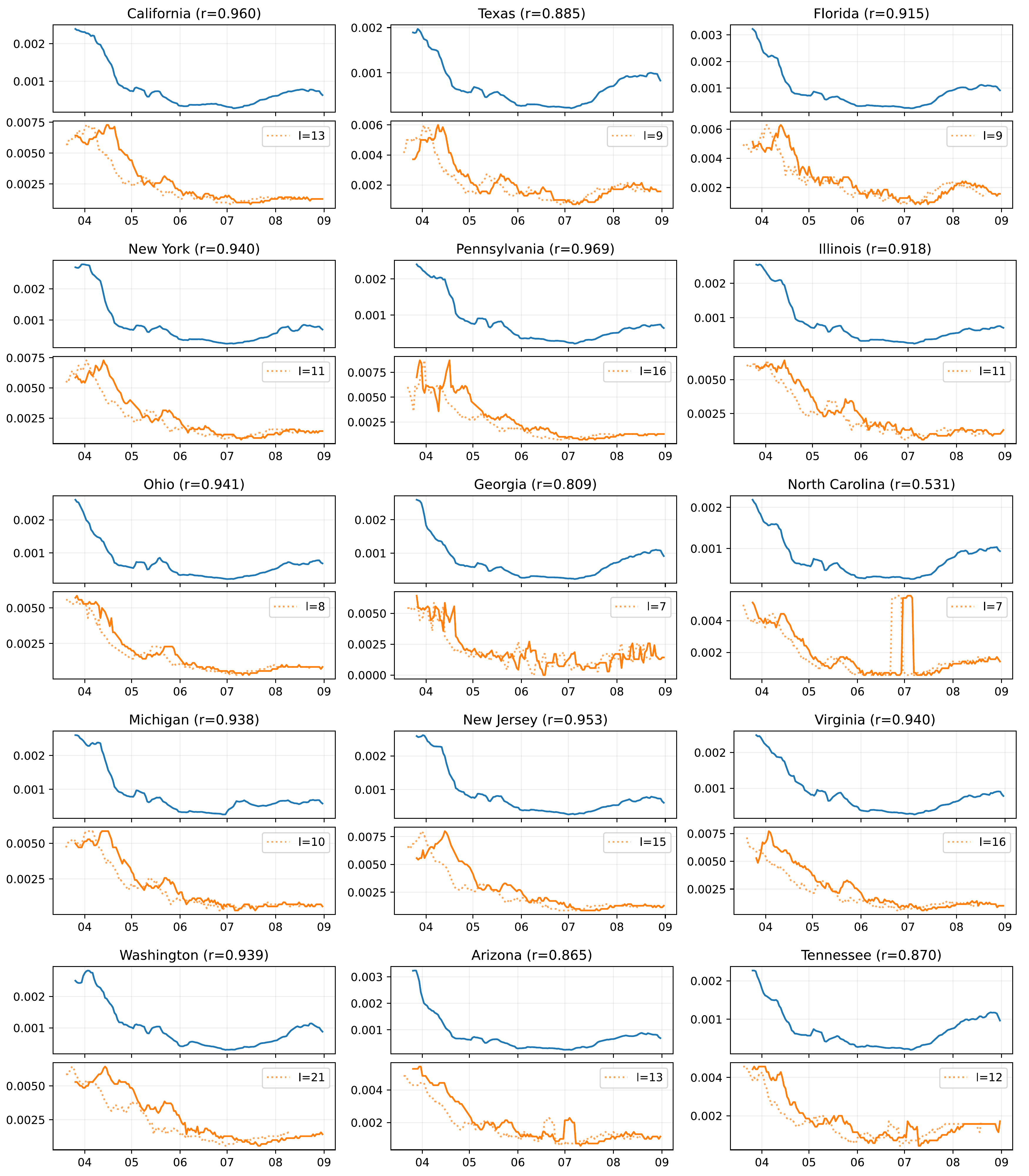}
    \caption{Comparing the vaccine intent time series (top, blue) to the CDC vaccination time series (bottom, orange) for the 15 largest states in the US. We compute the maximum correlation between the time series (reported in each subfigure's title) when the CDC time series is allowed to lag by $l$ days (reported in each legend).
    The dotted orange line in each bottom plot represents the adjusted CDC time series with the lag subtracted, so that the alignment between the time series can be more directly compared.}
    \label{fig:vi-cdc-over-time}
\end{figure}

\paragraph{Vaccination rates across counties.}
We also compare to CDC county-level vaccination rates. CDC county-level data are imperfect; for example, we find that even by the end of August 2021, 248 counties report less than 1\% of the population fully vaccinated, which is unrealistic given that over 60\% of the US population was vaccinated by this time.
We also find 585 counties where the ``completeness percent'', i.e., the percent of vaccination records that
include county of residence, is less than 80\%.
Following prior work using these data,\cite{tolbert2021kff} we exclude these incomplete counties from our analysis (which also cover all of the counties with less than 1\% vaccinated).
On the remaining counties, we compare our estimated vaccine rates $\vi{c}$ per county $c$ to the CDC’s fully vaccinated rates, cumulative up to August 31, 2021.
\begin{figure}
    \centering
    \includegraphics[width=10cm]{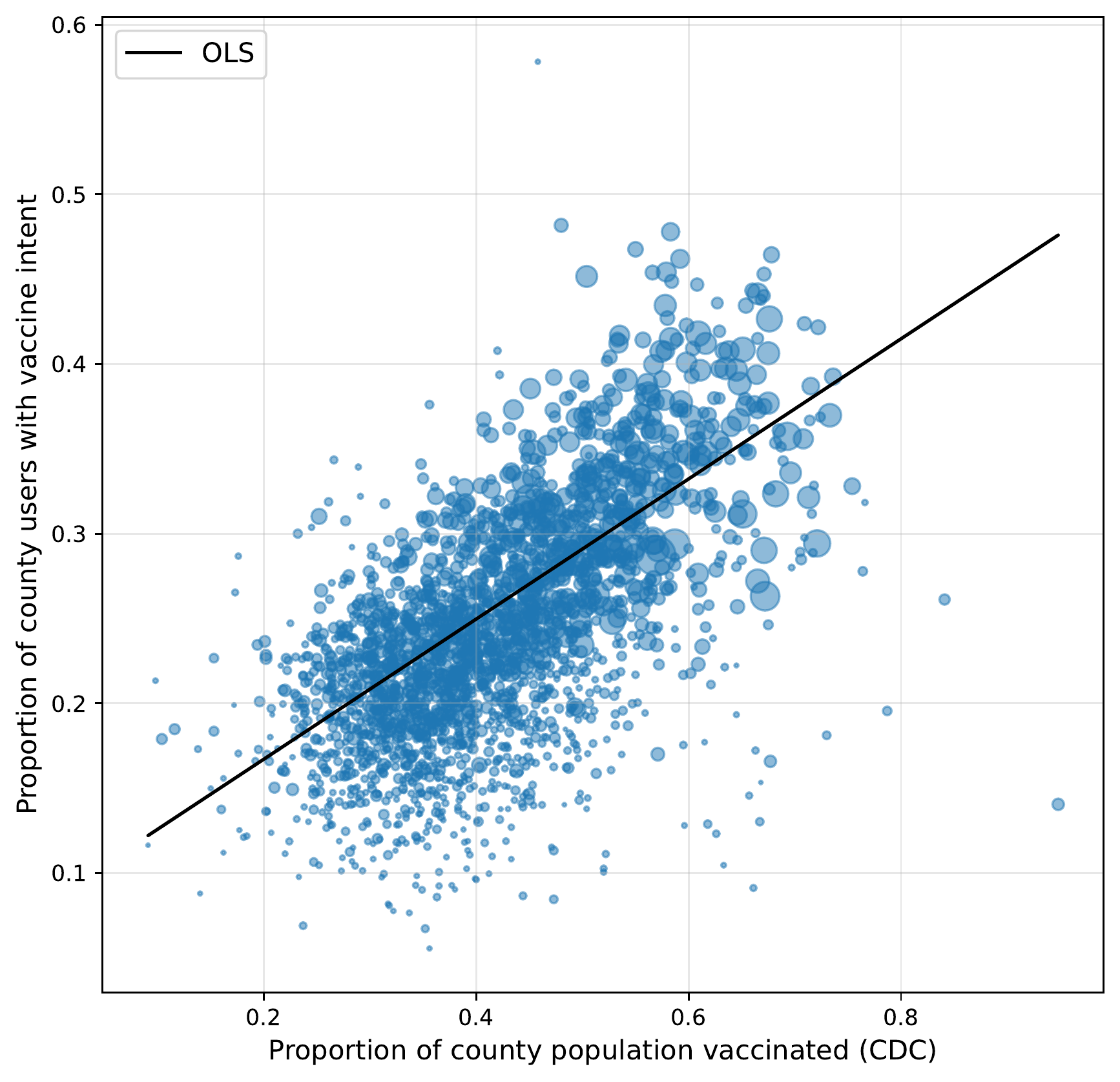}
    \caption{Comparing CDC county vaccination rates to estimated vaccine intent rates. The size of the dot and OLS fit are weighted by square root of county population.}
    \label{fig:cdc-county-cumul}
\end{figure}
The Pearson correlation, weighted by square root of county population, is $r = 0.68$ (Figure \ref{fig:cdc-county-cumul}), which is lower than the state-level correlation but still largely in agreement. 
Notably, we achieve higher correlations on counties where we expect higher-quality CDC reporting, which are counties with higher completeness percentages and larger populations, where reported proportions are less noisy.
If we remove the constraint on completeness percent, our correlation drops to $r = 0.54$. 
If we keep the completeness constraint at 80\% but remove weighting by population size, our correlation drops to $r = 0.58$. 
These observations suggest that discrepancies between our estimates and CDC data are at least in part driven by issues in CDC reporting, since our agreement improves on counties with higher-quality reporting.

\subsection{Comparison to Google search trends} \label{sec:methods-google}
\paragraph{Search trends over time.}
Following prior work using Bing data,\cite{suh2022info} we compare Bing and
Google queries to evaluate the representativeness of Bing search data. 
First, we compare daily search interest in the US over our studied time period from February 1 to August 31, 2021. 
Google Trends provides normalized search interest over time on Google, such that 100 represents the peak popularity for that time period, 50 means the term is half as popular, and 0 means ``there was not enough data for this term.''
To match this, for a given query, we compute the total number of times it was searched on Bing in the US per day, then we divide by the maximum number and multiply by 100. 
Again, we apply 1-week smoothing to both the Bing and Google time series.
We do not correct the Bing time series with Bing coverage here, since we cannot correct the Google time series with Google coverage, and we want the time series to be constructed as similarly as possible.

We evaluate 30 of the most common vaccine intent queries, including [cvs covid vaccine] and [covid vaccine finder].\footnote{We identify 30 representative vaccine intent queries from the top 100 vaccine intent queries, where we choose one standard query for each pharmacy that appears (e.g., [cvs covid vaccine]) and one for each location-seeking query (e.g., [covid vaccine near me]), and drop variants such as [cvs covid vaccines] and [covid 19 vaccine near me].}
\begin{figure}
    \centering
    \includegraphics[width=\linewidth]{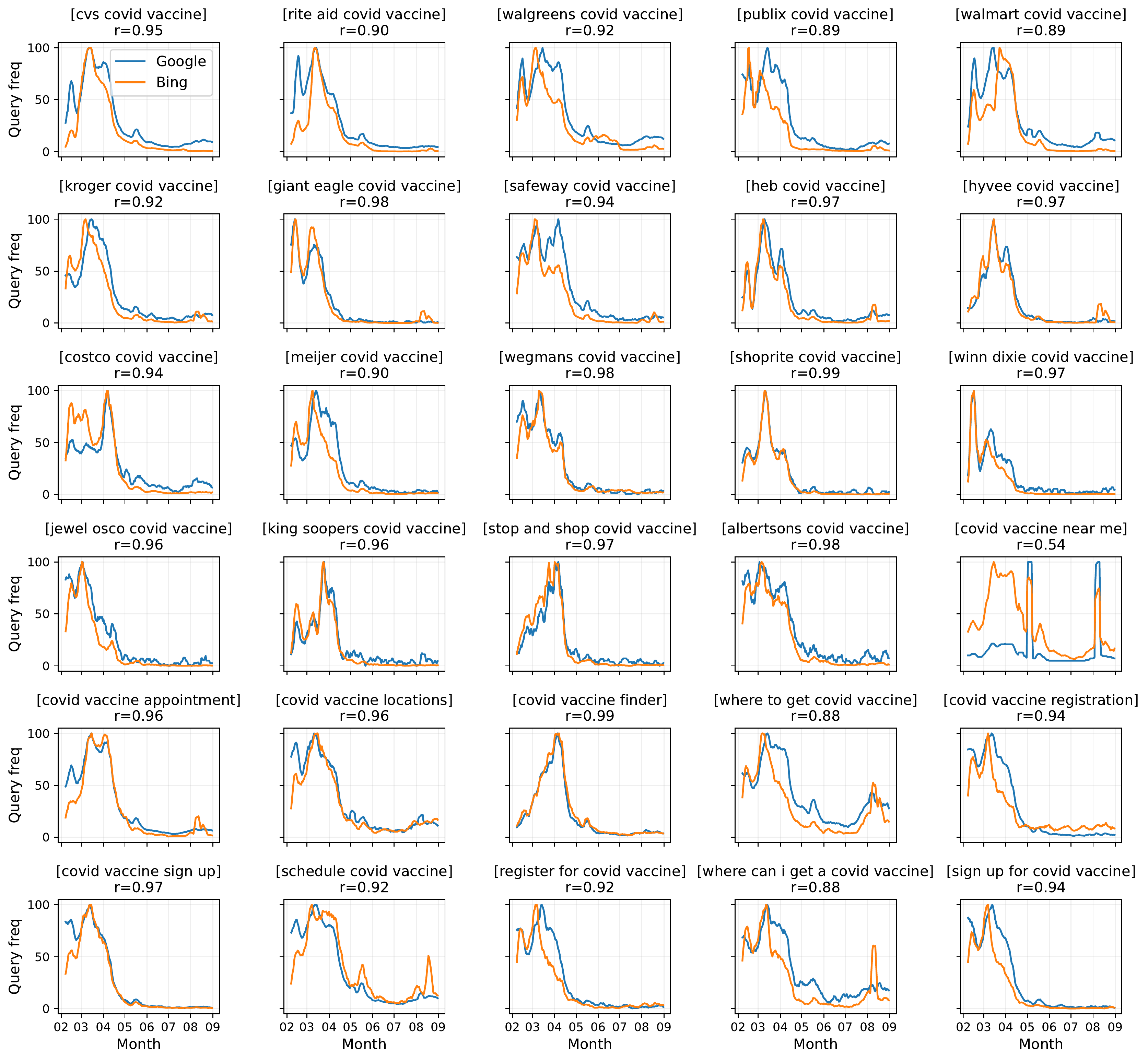}
    \caption{Comparing search trends over time on Google (blue) versus Bing (orange) for 30 of the most common vaccine intent queries. The y-axis represents query frequency, normalized so that 100 represents peak popularity in the US over the time period. 
    Pearson correlations between normalized trends are reported in each subfigure's title.}
    \label{fig:google-time}
\end{figure}
We observe strong Pearson correlations, with a median correlation of $r = 0.95$ (90\% CI, $0.88$-$0.99$) (Figure \ref{fig:google-time}). 
These correlations are similar to those reported by Suh et al. (2022)\cite{suh2022info}, who conduct an analogous longitudinal analysis comparing Bing and Google search trends on COVID-related
queries and report correlations from $r = 0.86$ to $0.98$. 
Remaining discrepancies between Bing and Google are likely due to differences in the populations using these search engines, as well as potential unreported details on how Google normalizes their search interest trends (e.g., Google may be normalizing differently for [covid vaccine near me], which shows unusual peaks in Google trends and is the the only query for which we do not observe a strong correlation).

\paragraph{Search trends across states.}
Google also provides normalized search interest across US states, where search interest is defined as the fraction of searches from that state that match the query and search interest is normalized across regions such that 100 represents maximum popularity. 
To imitate this process, we first assign each vaccine intent query to a state based on where the query originated. 
Then, we approximate the total number of queries (all queries, not just vaccine intent) from each state by summing over the query counts of the active users assigned to each state. 
We compute the fraction of queries from each state that match the query, then we divide by the maximum fraction and multiply by 100 to normalize across states. 
\begin{figure}
    \centering
    \includegraphics[width=\linewidth]{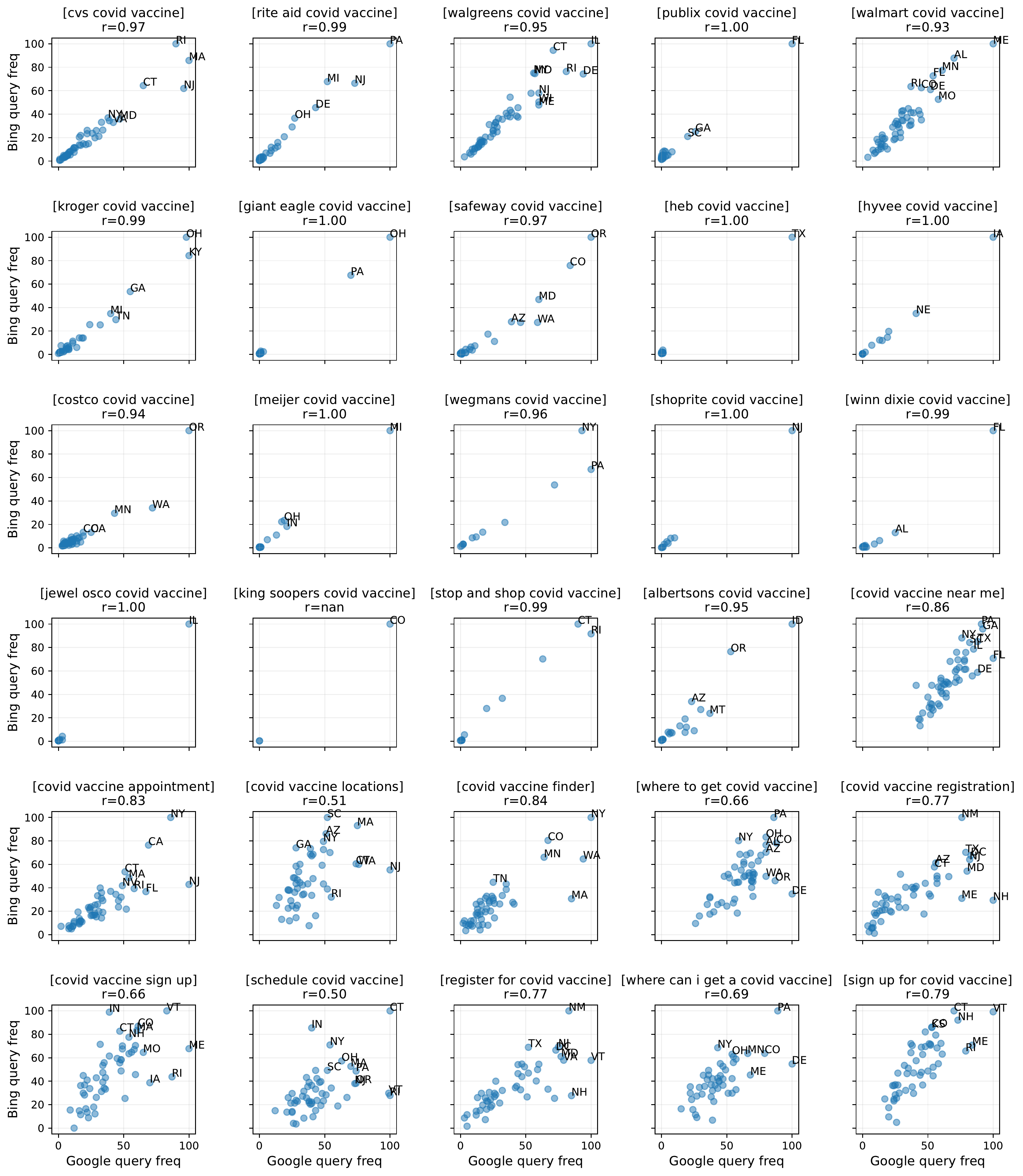}
    \caption{Comparing search trends across states on Google (x-axis) versus Bing (y-axis) for 30 of the most common vaccine intent queries. Following Google Trends, query frequency is measured as the fraction of the state's total queries that match this query, then normalized so that 100 corresponds to the maximum fraction over states.}
    \label{fig:google-states}
\end{figure}

We observe strong Pearson correlations in this analysis too, with a median correlation of $r = 0.95$ (90\% CI, $0.57$-$0.99$) across the same 30 vaccine intent queries (Figure \ref{fig:google-states}). 
The correlations tend to be stronger on the pharmacy-specific queries, where certain regions dominate, compared to general location-seeking queries such as [covid vaccine near me], which are trickier since they follow less obvious geographical patterns.
For the pharmacy-specific queries, we also observe substantial heterogeneity in terms of which region dominates.
For example, [publix covid vaccine] is more popular in southern states, with Florida exhibiting the maximum normalized search interest on Google (100), followed by Georgia (26) and South Carolina (20). 
Meanwhile, [cvs covid vaccine] is more popular in the Northeast, with the top states being Massachusetts (100), New Jersey (96), Rhode Island (90), and Connecticut (65). 
These differences, reflected in the Bing search trends too, once again highlight the need for regional awareness and representativeness when developing our vaccine intent classifier.
\section{Ontology of vaccine concerns on search} \label{sec:methods-ontology}
In this section, we describe how we match vaccine holdouts and vaccine early adopters (Section \ref{sec:methods-matching}), then how we construct a hierarchical ontology of vaccine concerns, based on the their clicks from April to August 2021 (Section \ref{sec:methods-ontology-process}).

\subsection{Identifying matched holdouts and early adopters} \label{sec:methods-matching}
Our study period covers search logs from February 1 to August 31, 2021.
First, we apply our vaccine intent classifier to the entire study period and identify 7.45M users who have expressed vaccine intent through queries and/or clicks.
Among those users, we define early adopters as those who showed their first vaccine intent before May (i.e., between February 1 and April 30, 2021) and vaccine holdouts as those who waited until July to show their first vaccine intent (i.e., between July 1 and August 31, 2021).
We choose these cutoffs since all US residents aged 16 and older were eligible for the vaccine by April 19,\cite{nyt2021eligibility} so those who waited until July to seek the vaccine were holding out.
Furthermore, to improve our ability to detect true holdouts, we require holdouts and early adopters to be active (i.e., issued at least 30 queries) in \textit{every} month during the study period, since if users were not active on Bing before July or August, their apparent lack of vaccine intent could be explained simply by low Bing usage during the earlier months.

To reduce potential confounding, we match each vaccine holdout to a unique vaccine early adopter from the same county, since we know that the populations seeking vaccination changed over time (Figures \ref{fig:classifier}d and \ref{fig:demo-quartile}b) and we do not want our comparisons to be overpowered by regional or demographic differences.
We also match on Bing usage, by requiring that the holdout's and early adopter's average monthly query counts during the study period do not differ by more than 10 queries.
We match on query count since our estimated time of first vaccine intent may be more delayed for individuals with less frequent Bing usage, so being labeled a holdout may be correlated with using Bing less frequently, which may reflect latent variables that also affect individuals' search behaviors.
To implement matching, we construct a bipartite graph between holdouts and early adopters, where an edge between a holdout and early adopter exists if that early adopter is a valid match for the holdout (they are from the same county and their average query counts are within 10 of each other).
Then, we run the Hopcroft–Karp algorithm on this graph, which finds the maximum matching, i.e., the largest set of edges where no two edges share an endpoint.\cite{hopcroft1971matching}
Since the number of early adopters greatly outnumbers the number of holdouts, we are able to match 98\% of our holdouts using this approach, resulting in 212,283 matched pairs.

\subsection{Constructing an ontology of search concerns} \label{sec:methods-ontology-process}
To analyze the vaccine concerns of holdouts and early adopters, our goal is to organize the vaccine-related URLs that they click on into a rich ontology.
We focus on their clicks from April to August 2021, since from April to June we can compare their vaccine concerns during the period while holdouts were eligible for the vaccine but still holding out, and from July to August, we can study how holdouts' concerns change as they finally express vaccine intent for the first time.
To construct our ontology, we combine computational and manual approaches: first, we use machine learning techniques to partition the URLs into clusters, and then we manually label each cluster and develop our three-tiered ontology of categories, subcategories, and URL clusters.

\paragraph{Automatically partitioning URLs into clusters.}
We begin by gathering all vaccine-related URLs (i.e., containing ``vaccin'' or ``vax'') that were clicked on by holdouts or early adopters from April to August 2021.
We drop all of the vaccine intent URLs, including the manual ones labeled by AMT or by regular expressions (Section \ref{sec:methods-amt}) and the ones discovered by GNNs (Section \ref{sec:methods-gnn}).
We also drop internal Microsoft links (e.g., ads) and URLs that were clicked on by fewer than 5 users (summed over holdouts and early adopters). 
After filtering, we are left with 32,811 vaccine-related URLs.
We construct the query-click graph for these URLs, based on the queries and clicks of holdouts and early adopters from April to August 2021. 
Since we only want to partition URLs, we then collapse the query-click graph to construct the URL co-click graph, where a weighted edge between two URLs indicates how connected these two URLs are by common queries.
Then, we partition the URLs into clusters by running the Louvain community detection algorithm\cite{blondel2008louvain} on the co-click graph.
Empirically, we find that clusters of 100-500 URLs are the most useful: they are large enough to represent substantial topics (as opposed to, for example, individual news stories), but not so large that their focus is unclear. 
We tune the Louvain resolution parameter $\gamma$, which controls how much the algorithm favors larger as opposed to smaller communities, to maximize the number of clusters within this range. 
We find that $\gamma=17$ allows us to achieve the largest number of clusters of size 100-500.

\paragraph{Manually constructing our ontology.}
We begin by labeling the clusters with at least 100 URLs, of which there are 79 when $\gamma=17$.
Each author labels the clusters independently, viewing a uniform random sample of 30 URLs from each cluster.
To aid our labeling process, we also view the 10 most frequent queries for each cluster, which we obtain by summing over all queries that led to clicks on URLs in the cluster.
When labeling, each author writes a free-text description of each cluster and marks whether it is clear.
We find that our automatic approach produces remarkably coherent topics and that the majority of clusters are clear.
\begin{table}
    \small
    \centering
    \begin{tabular}{p{0.8cm}|p{2cm}|p{11cm}|p{0.7cm}}
        \textbf{\# URLs} & \textbf{Top query} & \textbf{Top URLs} & \textbf{\% Clicks} \\
    \hline 
        206 & [cdc mask guidelines] & \url{https://www.cbsnews.com/news/cdc-mask-guidelines-covid-vaccine} & 8.0 \\
        & & \url{https://www.cdc.gov/media/releases/2021/p0308-vaccinated-guidelines.html} & 6.9 \\
        & & \url{https://www.usatoday.com/story/news/health/2021/05/13/covid-vaccine-cdc-variant-fda-clots-world-health-organization/5066504001} & 4.5 \\
        & & \url{https://www.nytimes.com/2021/05/13/us/cdc-mask-guidelines-vaccinated.html} & 4.4 \\
    \hline 
        139 & [vaers database & \url{https://www.cdc.gov/vaccinesafety/ensuringsafety/monitoring/vaers/index.html} & 17.0 \\
        & covid-19] & \url{https://rightsfreedoms.wordpress.com/2021/07/22/vaers-whistleblower-45000-dead-from-covid-19-vaccines-within-3-days-of-vaccination-sparks-lawsuit-against-federal-government} & 6.8 \\
        & & \url{https://www.theburningplatform.com/2021/07/03/latest-cdc-vaers-data-show-reported-injuries-surpass-400000-following-covid-vaccines} & 5.7 \\
        & & \url{https://vaersanalysis.info/2021/08/20/vaers-summary-for-covid-19-vaccines-through-8-13-2021} & 4.9 \\
    \hline 
        137 & [religious exemption & \url{https://www.verywellfamily.com/religious-exemptions-to-vaccines-2633702} & 16.5 \\
        & for covid-19 vaccination] & \url{https://www.fisherphillips.com/news-insights/religious-objections-to-mandated-covid-19-vaccines-considerations-for-employers.html} & 5.1 \\
        & & \url{https://www.law360.com/articles/1312230/employers-should-plan-for-vaccine-religious-exemptions} & 3.9 \\
        & & \url{https://www.kxly.com/who-qualifies-for-a-religious-exemption-from-the-covid-19-vaccine} & 3.3 \\
    \hline 
        113 & [johnson and johnson & \url{https://www.openaccessgovernment.org/side-effects-johnson-johnson-vaccine/109505} & 20.3 \\
        & side effects] & \url{https://www.healthline.com/health/vaccinations/immunization-complications} & 8.1 \\
        & & \url{https://www.msn.com/en-us/health/medical/these-are-the-side-effects-from-the-johnson-and-johnson-covid-19-vaccine/ar-bb1f03fq} & 4.3 \\
        & & \url{https://www.healthline.com/health-news/mild-vs-severe-side-effects-from-the-johnson-and-johnson-covid-19-vaccine-what-to-know} & 4.3 \\
    \hline 
        
    \end{tabular}
    \caption{The 4 highest-modularity clusters with at least 100 URLs. For each cluster, we provide its number of URLs, its most frequent query, its top 4 URLs (by click frequency), and percentage of clicks over all clicks on URLs in the cluster that the URL accounts for.}
    \label{tab:cluster-samples}
\end{table}
In Table \ref{tab:cluster-samples}, we provide a sample of clusters.
From the top query and most frequently clicked URLs, we observe distinct topics covered in each cluster: one on CDC masking guidelines after vaccination, one on the Vaccine Adverse Event Reporting System (VAERS),\cite{vaers} one about religious exemptions for COVID-19 vaccine requirements, and one about side effects of the Johnson \& Johnson vaccine.

Based on our descriptions of the clusters, we identify 8 top categories and 36 subcategories of vaccine concerns (Figure \ref{fig:ontology}).
For example, under Vaccine Safety, we include the subcategories of normal side effects (e.g., sore arms), severe side effects (e.g., blood clots), concerns about reproductive health, fear of vaccine-caused deaths, ``eerie'' fears (e.g., myths about vaccine shedding or becoming magnetic\cite{cdc2023myths}), vaccine development (e.g., pace of development, ingredients in the vaccine), and FDA approval.
As we show in the following section, these fine-grained subcategories allow us to study nuances in vaccine concerns; for example, holdouts and early adopters are both concerned about vaccine safety, but focus on different aspects of it.
Finally, we take a second pass through the URL clusters to sort them into subcategories, allowing each cluster to belong to at most 2 subcategories.
For example, for the cluster about religious exemptions (Table \ref{tab:cluster-samples}), we sort it into both the religious concerns and exemptions subcategories.
During this second pass, we label all clusters with at least 30 URLs. 
For the clusters that are unclear, we rerun Louvain community detection on the cluster (with the default $\gamma=1$, since the number of URLs being partitioned is much smaller) to try to identify smaller groups of URLs that we can assign to subcategories.
We are able to assign most clusters to subcategories, but there are some that we leave out, since they cover miscellaneous topics such as one-off news stories or specific interests (e.g., how to store vaccines) that do not clearly belong to any of our subcategories.
At the end of our process, our constructed ontology consists of 24,726 URLs (75\% of all 32,811 vaccine-related URLs).

\section{Main and supplemental analyses} \label{sec:methods-analysis}
In this section, we provide additional details for the analyses presented in the main text and describe supplemental analyses that we conducted.

\subsection{Granular trends in vaccine seeking}
In Figure~\ref{fig:vi-trends}, we visualize our estimates of ZCTA vaccine intent rates.
For privacy reasons, we focus our analyses on ZCTAs where the number of active Bing users $N(b, z) \geq 50$ and the Census population size $N(z) \geq 50$, keeping around 21,000 ZCTAs and covering 97\% of the US population. 
We estimate coverage-corrected vaccine intent rates $\vi{z}$ per ZCTA $z$ following Eq.~\ref{eqn:vi-z}.

\paragraph{Map visualizations.}
For our map visualizations, we use ZCTA, county, and state shapefiles from the 2020 US Census.\cite{census_shapefiles}
In Figure~\ref{fig:vi-trends}a, we visualize vaccine intent rates for the entire US.
Since we cannot estimate vaccine intent rates for all ZCTAs, first we visualize vaccine intent rates per county (following Eq.~\ref{eqn:vi-set}), then we overlay ZCTA vaccine intent rates.
Additionally, we overlay the boundaries of the states in white, to emphasize state-level trends as well as  reveal substantial heterogeneity in vaccine intent rates even within the same state.
To explore this heterogeneity in greater detail, in Figure~\ref{fig:vi-trends}b, we zoom in on the five counties in New York City, corresponding to the five boroughs: Manhattan (New York County), Queens (Queens County), Brooklyn (Kings County), Bronx (Bronx County), and Staten Island (Richmond County).
ZCTAs are well-represented here, given the population of the city, so we only visualize ZCTA vaccine intent rates without county-level rates.
Also, instead of drawing state boundaries in white, we draw county boundaries in white, to emphasize trends per county and heterogeneity within each county.
For example, we see that Manhattan and Queens have higher estimated rates of vaccine intent, and within Queens, ZCTAs in the northern half have higher rates, aligning with reported local vaccination rates in New York City earlier in the pandemic.\cite{nyt2021nyc}

\paragraph{Measuring demographic trends.}
To characterize this heterogeneity, we compare ZCTA vaccine intent rates to demographic variables.
First, we measure the Pearson correlation between vaccine intent rate and each demographic variable, with correlation weighted by the square root of the ZCTA population.
In Figure~\ref{fig:vi-trends}c, we plot the results, ordered by strongest positive to strongest negative correlation, with 95\% confidence intervals.\cite{altman1988cis}
Our results agree with prior literature, finding positive correlations with percent with Bachelor degree, median income, population per square meter, percent 65 and over, percent Asian, percent White, and percent female, and negative correlations with percent Republican, percent under 18, percent Black, and percent Hispanic.\cite{kreps2020factors,troiano2021hesitancy,joshi2021predictors,yasmin2021review}

To investigate more granular demographic trends, we measure correlations per state (only including the ZCTAs in the state) for the 10 largest states in the US.
For this finer-grained analysis, we drop percent Republican, since we only have vote share at the county-level, but we keep all other demographic variables, which we have per ZCTA.
We find that correlations are mostly consistent in sign across states, but the magnitude differs significantly (Figure~\ref{fig:state-demo}).
For example, the positive correlation with percent 65 and over is around 2x as high in Florida as it is in the second highest states, reflecting the large senior population in Florida and the push for seniors to get vaccinated.
In most states, we also see positive correlations for percent Asian and percent White, and negative correlations for percent Black and percent Hispanic, aligning with prior research on racial and ethnic disparities in COVID-19 vaccination rates.\cite{siegel2022disparities}
Positive and negative correlations for race are particularly strong in certain states, including New York and Florida for percent White/Black, and California and New York for percent Hispanic.
\begin{figure}
    \centering
    \includegraphics[width=\linewidth]{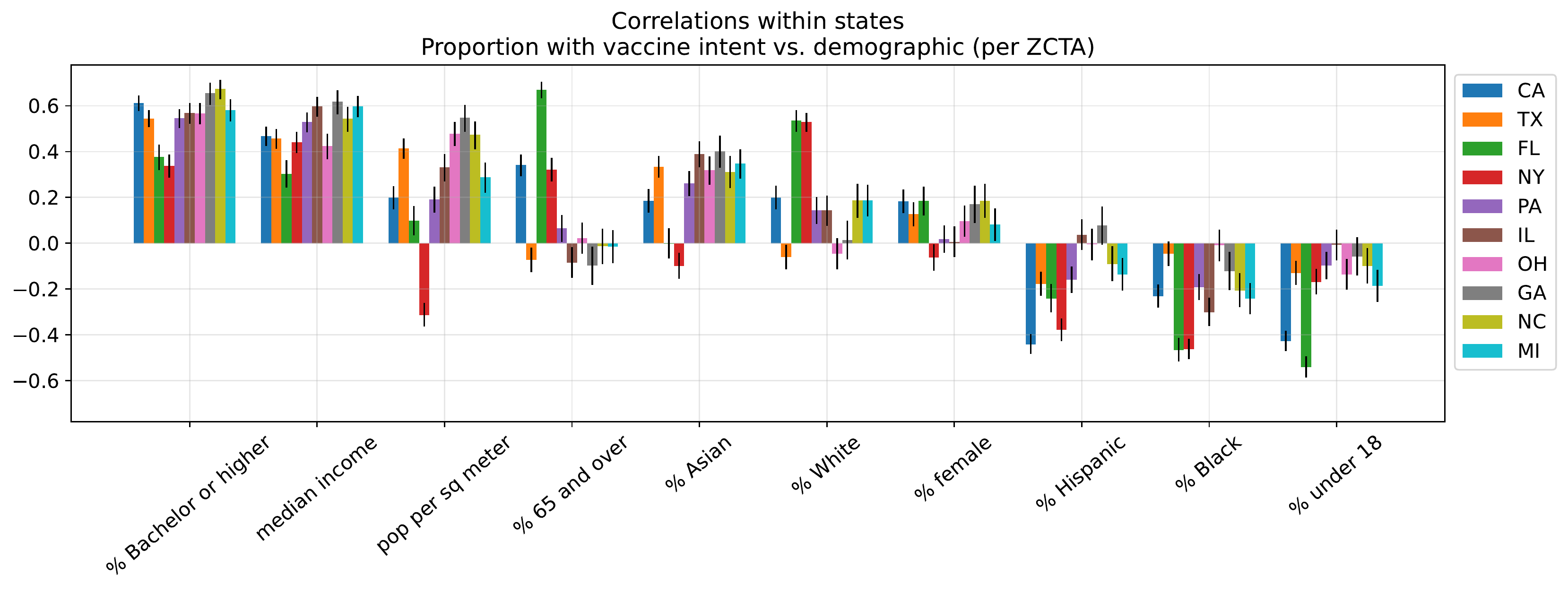}
    \caption{Correlations between ZCTA vaccine intent rate and demographic variables, for the 10 largest US states. Error bars indicate 95\% CIs.}
    \label{fig:state-demo}
\end{figure}

We also conduct a supplementary analysis where we use a different statistic to characterize demographic trends.
We refer to this statistic as a ``quartile comparison'': we separate ZCTAs into top and bottom quartiles based on a demographic variable, then compare the quartiles' average vaccine intent rates (weighted by population size, following Eq.~\ref{eqn:vi-set}).
Note that we compute quartile cutoffs based on \textit{all} ZCTAs in the US Census, not only the ZCTAs that we are able to keep in our study.
We compute bootstrapped CIs for these ratios by resampling the ZCTAs included (with replacement) then resampling each ZCTA's vaccine intent rate by sampling its number of users with vaccine intent from Binom($N(b,z)$, $\tilde{p}(v,z)$). 
We generate 1000 bootstrapped samples and, from the resulting distribution of ratios, report the 2.5th and 97.5th percentiles as the 95\% CI.
We find that demographic trends measured by quartile comparison (Figure~\ref{fig:demo-quartile}a) largely agree with what we saw with correlations: population per square meter, percent with Bachelor degree, and median income continue to have the strongest positive relationships with vaccine intent, followed by percent Asian, percent 65 and over, and percent female. Percent Republican has the strongest negative relationship with vaccine intent, followed by percent under 18; the only differences with correlation are in percent White, Black, and Hispanic.
Separating ZCTAs into top and bottom quartiles enables us to compute vaccine intent rates per quartile over time. We compute a time series per quartile in the same way that we computed the vaccine intent rate per state over time in Section~\ref{sec:methods-cdc}, by estimating the proportion of users who showed their \textit{first} vaccine intent per day.
Then, comparing the ratio of time series reveals changes in demographic trends over time (Figure~\ref{fig:demo-quartile}b). 
For example, we estimate that older ZCTAs were much likelier to seek the vaccine early in 2021 but this trend fell over time, reflecting how the US vaccine rollout first prioritized seniors then expanded to general eligibility,\cite{cnbc2021seniors,nyt2021eligibility}
and we see an increase in vaccine intent from more Republican ZCTAs in summer 2021, reflecting new calls from Republican leaders to get vaccinated\cite{wapo2021gop} and a self-reported uptick in vaccinations among Republicans.\cite{gallup2021sept}
\begin{figure}
    \centering
    \includegraphics[width=\linewidth]{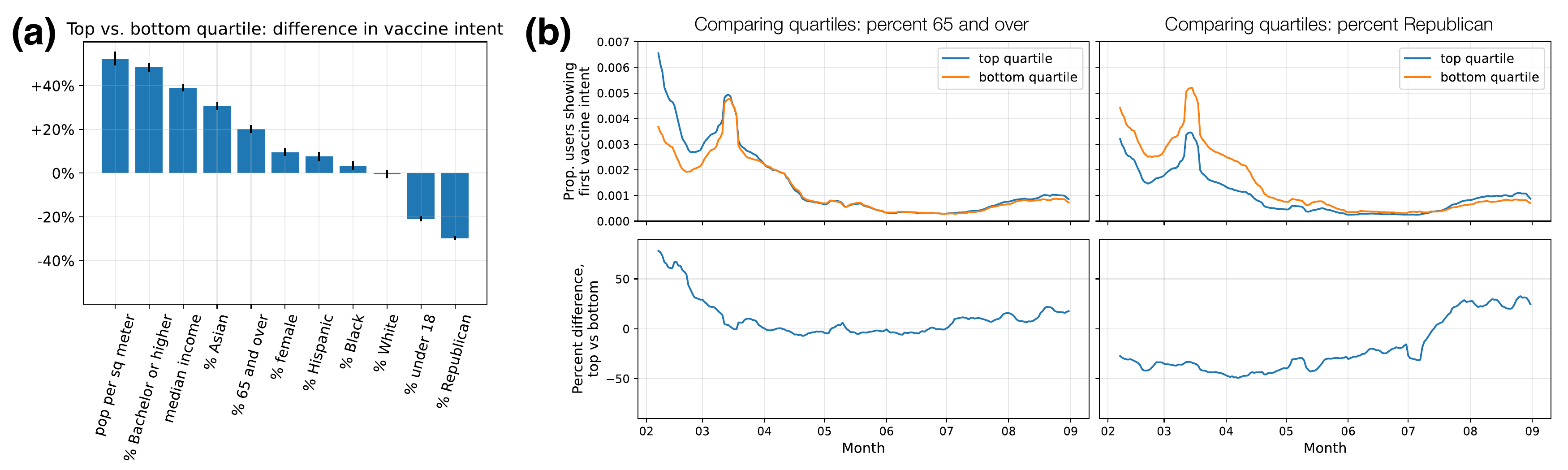}
    \caption{Demographic trends in vaccine intent, measured by quartile comparison. \textbf{(a)} We separate ZCTAs into top and bottom quartiles based on each demographic variable and compute the percent difference in the top quartile's vs. bottom quartile's average vaccine intent rate. Error bars indicate bootstrapped 95\% CIs. \textbf{(b)} We can quantify changes over time in demographic trends by estimating average vaccine intent rates per quartile over time (top) and computing their percent difference (bottom).}
    \label{fig:demo-quartile}
\end{figure}

\subsection{Analyses of vaccine concerns and news consumption} \label{sec:methods-concerns}
\paragraph{Click ratios comparing two groups.}
In several analyses, we compare click probabilities for two groups, such as the probability that a holdout versus an early adopter clicks on untrusted news.
For each group $g$, such as holdouts, first we gather all their relevant clicks $R_g$ (e.g., news-related, vaccine-related) from the time period of the analysis.
Then, we identify the ``positive'' subset of clicks $S_g \subseteq R_g$, such as clicks on untrusted news or a specific vaccine subcategory.
Then, we compute $p_g$, the weighted average probability over clicks that the click is in the positive set, weighted by the user's ZCTA's inverse Bing coverage (Section~\ref{sec:methods-bing-coverage}):
\begin{align}
    p_g = \frac{\sum_{x_i \in R_g} \mathbbm{1}[x_i \in S_g] \frac{N(z_i)}{N(b, z_i)}}{\sum_{x_j \in R_g} \frac{N(z_j)}{N(b, z_j)}}, \label{eqn:pg}
\end{align}
where $z_i$ represents the ZCTA of click $x_i$'s user.
To compare two groups, such as holdouts and early adopters, we take the ratios of their click probabilities.
We compute bootstrapped CIs for these ratios by repeatedly resampling the clicks in each group (with replacement), recomputing the group averages, and recomputing the ratio between group averages.
Then, from the resulting distribution of ratios, computed over 1000 bootstrapped samples, we report the 2.5th and 97.5th percentiles as the 95\% CI.

We compare to an alternative measure that computes the fraction of clicks in the positive set \textit{per user}, then computes the weighted average over users, weighting again by their ZCTA's inverse Bing coverage. 
We find in practice that the measures result in very similar trends ($r > 0.95$), which is expected, since we have over 200,000 users per group and most users have clicks (since we only keep users who are active in every month during our study period), so our measure is not dominated by a few users' clicks even if we do not compute user-specific fractions first.
So, we use our measure $p_g$ for simplicity, which is more straightforward to compute and bootstrap.

\paragraph{News consumption of holdouts versus early adopters.}
In this analysis, we compare the news consumption of holdouts versus matched early adopters from April to June 2021. 
We use labels from Newsguard, which include 7,226 news domains and 2,744 untrusted domains.
For each group $g$, the relevant set of clicks $R_g$ includes all clicks on news domains.
First, we compare the probabilities of clicking on untrusted news, so we set positive clicks $S_g$ to clicks on untrusted news domains, and compute average probabilities for holdouts and early adopters according to Eq.~\ref{eqn:pg}. 
We find that holdouts are 69\% (95\% CI, 67\%-70\%) more likely to click on untrusted news, compared to their matched early adopters.
To compute click ratios for a specific news domain, we set $S_g$ to clicks on that domain, and compute weighted probabilities and ratios accordingly. 
We compute and visualize ratios for domains that receive at least 0.0001\% of the total news clicks, which leaves 3,282 domains in Figure~\ref{fig:concerns}a.

\paragraph{Vaccine concerns of holdouts versus early adopters.}
In this analysis, we compare the vaccine concerns of holdouts versus matched early adopters from April to June 2021.
Our relevant set of clicks $R_g$ now includes each group's vaccine-related clicks (i.e., containing ``vaccin'' or ``vax'') during this time period. 
To analyze click patterns for a given category of concerns (or subcategory), we set positive clicks $S_g$ to clicks on URLs belonging to that category, according to our ontology, and compute probabilities and ratios accordingly (Eq.~\ref{eqn:pg}).
In Figure~\ref{fig:concerns}b, we visualize holdouts' click proportions over time for the 6 main categories: Vaccine Requirements, Community, Effectiveness, Safety, Incentives, and Information.
\begin{table}[]
    \centering
    \begin{tabular}{c|c|c|c|c}
        \textbf{Category} & \textbf{Subcategory} & \textbf{Ratio} & \textbf{2.5th pctl} & \textbf{97.5th pctl} \\
        \hline 
            Requirements & -- & 1.58 & 1.55 & 1.63 \\
            & Exemption & 2.06 & 1.72 & 2.48 \\
            & Anti-mandate & 2.00 & 1.93 & 2.09 \\
            & Fake proof & 1.98 & 1.74 & 2.24 \\ 
            & Travel & 1.40 & 1.34 & 1.47 \\
            & Proof & 1.20 & 1.14 & 1.28 \\
            & Employment & 1.16 & 1.02 & 1.30 \\
        \hline
            Community & -- & 1.24 & 1.20 & 1.27 \\
            & Religious concerns & 2.50 & 2.24 & 2.80 \\
            & Expert anti-vax & 2.45 & 2.30 & 2.61 \\
            & Figure anti-vax & 2.23 & 2.01 & 2.48 \\ 
            & News on hesitancy & 1.20 & 1.13 & 1.26 \\ 
            & Vaccine rates & 0.59 & 0.56 & 0.62 \\
        \hline
            Safety & -- & 1.01 & 0.99 & 1.02 \\ 
            & Eerie fears & 2.11 & 2.03 & 2.19 \\ 
            & FDA approval & 1.72 & 1.58 & 1.87 \\ 
            & Vaccine-caused deaths & 1.67 & 1.61 & 1.74 \\
            & Vaccine development & 1.54 & 1.49 & 1.60 \\
            & Reproductive health & 1.27 & 1.19 & 1.35 \\
            & Severe side effects & 0.88 & 0.85 & 0.91 \\ 
            & Normal side effects & 0.42 & 0.41 & 0.43 \\
        \hline
            Effectiveness & -- & 0.84 & 0.82 & 0.87 \\ 
            & Natural immunity & 1.30 & 1.21 & 1.39 \\ 
            & Efficacy against variants & 1.24 & 1.13 & 1.37 \\ 
            & Breakthrough cases & 1.10 & 1.05 & 1.16 \\
            & Efficacy from studies & 0.57 & 0.55 & 0.60 \\
        \hline 
            Incentives & -- & 0.79 & 0.75 & 0.82 \\ 
        \hline 
            Information & -- & 0.72 & 0.71 & 0.74 \\
            & Decision-making & 1.13 & 1.06 & 1.20 \\ 
            & Post-vax guidelines & 1.10 & 1.05 & 1.14 \\ 
            & Special populations & 0.87 & 0.81 & 0.94 \\ 
            & Johnson \& Johnson & 0.56 & 0.52 & 0.59 \\ 
            & Pfizer & 0.53 & 0.50 & 0.57 \\
            & Comparison & 0.53 & 0.48 & 0.57 \\ 
            & Moderna & 0.37 & 0.35 & 0.39 \\ 
        \hline 
            Availability & -- & 0.22 & 0.21 & 0.24 \\ 
            & Boosters & 0.78 & 0.65 & 0.91 \\ 
            & Children & 0.70 & 0.64 & 0.76 \\ 
            & Locations & 0.05 & 0.04 & 0.06 \\
        \hline
    \end{tabular}
    \caption{Vaccine concerns of holdouts vs. early adopters. We report ratios of click probabilities, with bootstrapped 95\% CIs, within vaccine-related clicks from April to June 2021. A higher ratio indicates a greater lean towards holdouts; a lower ratio indicates a greater lean towards early adopters. Categories are ordered from highest to lowest ratio and subcategories within categories are also ordered from highest to lowest ratio.}
    \label{tab:holdout-ea-ratios}
\end{table}
We report click ratios and 95\% CIs for all categories and subcategories in Table~\ref{tab:holdout-ea-ratios}, and visualize subcategory click ratios in Figure~\ref{fig:concerns}c.
In both Figures~\ref{fig:concerns}b and~\ref{fig:concerns}c, categories/subcategories are ordered top to bottom and colored from yellow to dark purple in terms of most holdout-leaning to most early adopter-leaning.

We conduct an additional analysis to analyze variation in holdout concerns across demographic groups. 
For a given demographic variable, we compute its median value across all ZCTAs, split holdouts into those from ZCTAs above the median versus those from ZCTAs below the median, then compare the vaccine concerns of those two groups of holdouts (by measuring their click ratios).
We find significant variability across demographic groups in terms of holdout concerns (Figure~\ref{fig:concerns-demo}).
Compared to holdouts from more Republican-leaning ZCTAs, holdouts from more Democrat-leaning ZCTAs were far more interested in requirements around employee mandates and vaccine proof, which may be because jurisdictions run by Democrats were likelier to have vaccine requirements,\cite{cnn2021nyc,deadline2021la} while several Republican governors in fact banned such requirements.
Meanwhile, holdouts from more Republican-leaning ZCTAs were more interested in eerie vaccine fears, fears of vaccine-caused deaths, and vaccine incentives.
We also find that, compared to holdouts from lower-income ZCTAs, holdouts from higher-income ZCTAs were significantly more interested in vaccine requirements, vaccine rates, and anti-vaccine messages from experts and high-profile figures, while holdouts from lower-income ZCTAs were more interested in vaccine incentives and religious concerns about the vaccine.

\begin{figure}
    \centering
    \includegraphics[width=14.5cm]{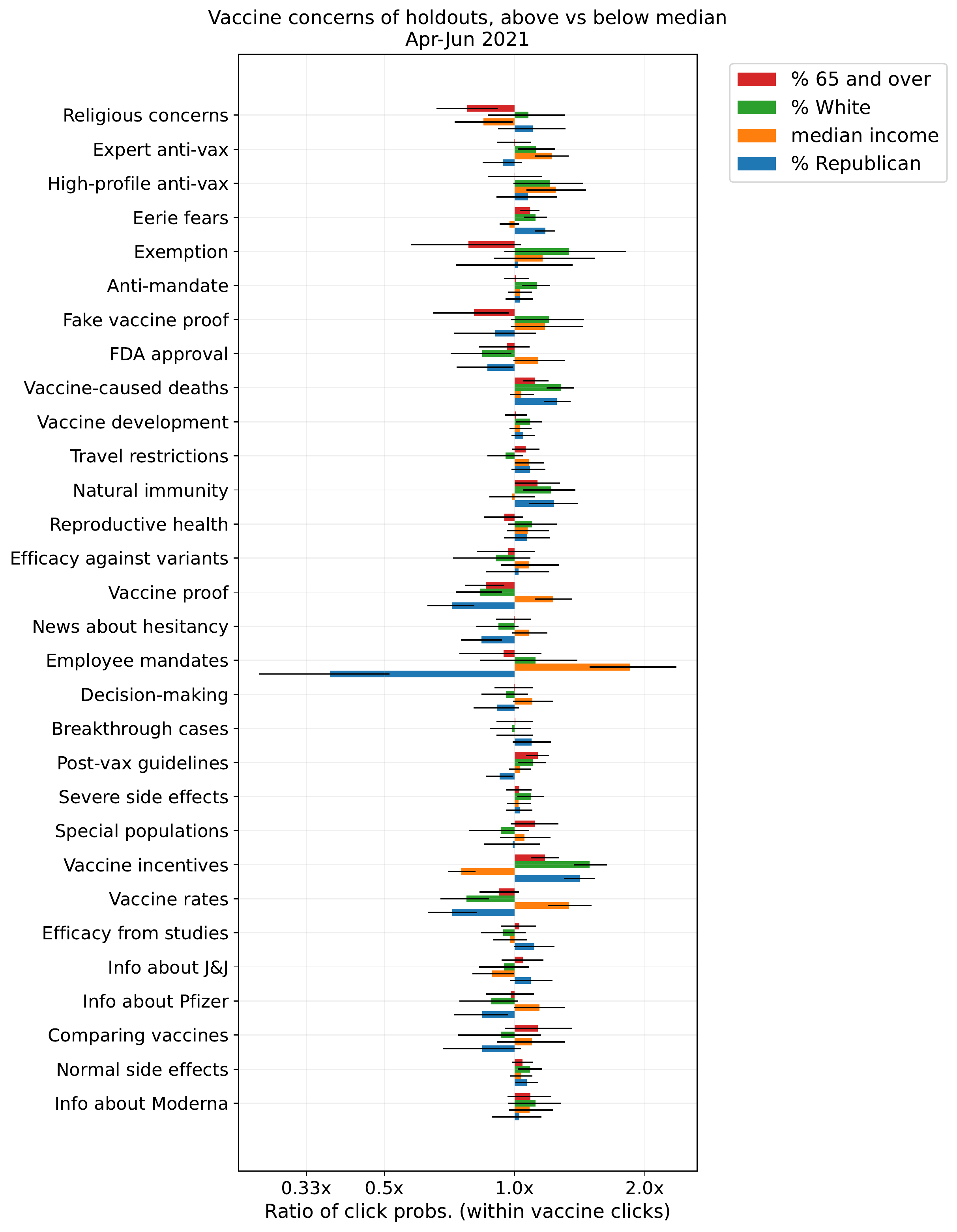}
    \caption{Variability in holdout concerns across demographic groups. For each demographic variable (e.g., percent Republican), we compare the concerns of holdouts from ZCTAs above the variable's median versus holdouts from ZCTAs below the median. Subcategories are ordered from most holdout-leaning to most early adopter-leaning, following Figure~\ref{fig:concerns}c. Error bars indicate bootstrapped 95\% CIs.}
    \label{fig:concerns-demo}
\end{figure}

\paragraph{Vaccine concerns of holdouts near vaccine intent versus not.}
In our final analysis, we analyze how holdouts' vaccine concerns change in the small window leading up to and following their expressed vaccine intent.
We split holdouts' vaccine-related clicks from July to August 2021 into two groups: clicks when the holdout is within 3 days (either before or after) of expressing vaccine intent and clicks outside of that range.
Since precision of vaccine intent timing is particularly important for this analysis, we focus on vaccine intent expressed either through a vaccine intent query or a manually labeled vaccine intent URL (either from AMT or regular expression).
Then, as before, we compute ratios per subcategory, where the set of positive clicks $S_g$ are those that match the subcategory and we compute probabilities per group according to Eq.~\ref{eqn:pg}.
\begin{table}[]
    \centering
    \begin{tabular}{c|c|c|c|c}
        \textbf{Category} & \textbf{Subcategory} & \textbf{Ratio} & \textbf{2.5th pctl} & \textbf{97.5th pctl} \\
        \hline 
            Availability & -- & 4.68 & 4.49 & 4.88 \\ 
            & Locations & 35.39 & 32.05 & 39.17 \\
            & Boosters & 2.02 & 1.85 & 2.20 \\ 
            & Children & 1.01 & 0.90 & 1.12 \\ 
        \hline 
            Incentives & -- & 2.46 & 2.30 & 2.61 \\ 
        \hline 
            Information & -- & 1.39 & 1.36 & 1.42 \\
            & Johnson \& Johnson & 4.29 & 4.07 & 4.54 \\ 
            & Comparison & 2.52 & 2.33 & 2.72 \\ 
            & Moderna & 1.20 & 1.10 & 1.29 \\ 
            & Special populations & 1.19 & 1.08 & 1.30 \\ 
            & Pfizer & 1.07 & 1.01 & 1.13 \\
            & Decision-making & 1.06 & 0.99 & 1.14 \\ 
            & Post-vax guidelines & 0.48 & 0.44 & 0.52 \\ 
        \hline
            Safety & -- & 0.82 & 0.81 & 0.84 \\ 
            & Normal side effects & 1.30 & 1.26 & 1.35 \\
            & Severe side effects & 1.06 & 1.02 & 1.11 \\ 
            & Reproductive health & 0.85 & 0.77 & 0.94 \\
            & FDA approval & 0.75 & 0.70 & 0.81 \\ 
            & Vaccine development & 0.68 & 0.64 & 0.71 \\
            & Eerie fears & 0.64 & 0.61 & 0.68 \\ 
            & Vaccine-caused deaths & 0.46 & 0.44 & 0.49 \\
        \hline
            Effectiveness & -- & 0.80 & 0.77 & 0.83 \\ 
            & Efficacy from studies & 1.20 & 1.13 & 1.28 \\
            & Efficacy against variants & 0.76 & 0.71 & 0.82 \\ 
            & Natural immunity & 0.67 & 0.62 & 0.72 \\ 
            & Breakthrough cases & 0.58 & 0.54 & 0.62 \\
        \hline
            Requirements & -- & 0.61 & 0.59 & 0.63 \\
            & Proof & 1.02 & 0.96 & 1.08 \\
            & Fake proof & 0.81 & 0.73 & 0.88 \\ 
            & Travel & 0.64 & 0.57 & 0.72 \\
            & Employment & 0.58 & 0.49 & 0.69 \\
            & Anti-mandate & 0.52 & 0.49 & 0.54 \\
            & Exemption & 0.36 & 0.33 & 0.40 \\
        \hline
            Community & -- & 0.57 & 0.54 & 0.59 \\
            & News on hesitancy & 0.86 & 0.80 & 0.92 \\ 
            & Vaccine rates & 0.59 & 0.55 & 0.64 \\
            & Expert anti-vax & 0.49 & 0.45 & 0.52 \\
            & Religious concerns & 0.40 & 0.36 & 0.44 \\
            & Figure anti-vax & 0.30 & 0.20 & 0.42 \\ 
        \hline 
    \end{tabular}
    \caption{Vaccine concerns of holdouts close to vaccine intent versus not. We report ratios of click probabilities, with bootstrapped 95\% CIs, within vaccine-related clicks from July to August 2021. A higher ratio indicates elevated interest near vaccine intent; a lower ratio indicates reduced interest. Categories are ordered from highest to lowest ratio and subcategories within categories are also ordered from highest to lowest ratio.}
    \label{tab:near-vi-ratios}
\end{table}
We report resulting ratios for all categories and subcategories in Table~\ref{tab:near-vi-ratios}, and visualize subcategory ratios in Figure~\ref{fig:concerns}d.
To facilitate comparison between Figures~\ref{fig:concerns}c and d, we keep the ordering of subcategories the same.
This design choice highlights how Figure~\ref{fig:concerns}d nearly reverses Figure~\ref{fig:concerns}c, meaning that near when holdouts express vaccine intent, their concerns become much more like the concerns of early adopters, with a few important differences.

We conduct a supplemental analysis to test the impact of controlling for time, which could be a confounder.
Specifically, from July 1 to August 31, 2021, holdouts became increasingly likely to express vaccine intent (Figure~\ref{fig:classifier}d), so comparing clicks within a small window of vaccine intent versus outside of that window may also capture changes in interest over time, separate from changes related to vaccine intent.
In our main analysis, we estimate the probabilities that a vaccine-related click from a holdout is on a given subcategory when the holdout is in their vaccine intent window versus not and compute the ratio of those probabilities.
We can also estimate these probabilities using a logistic regression model, where we fit a model to predict whether a vaccine-related click from a holdout is on a given subcategory, conditioned on whether the holdout is in their vaccine intent window.
Then, we can test the impact of controlling for time by extending the model to also condition on the date of the click.

For a click $x_i$, let $v_i \in \{0, 1\}$ represent whether the click is within 3 days of the user expressing vaccine intent and let $t_i \in \{\text{2021/07/01}, \cdots, \text{2021/08/31}\}$ represent the date of the click.
Then, for a given subcategory $s$, the full logistic regression model estimates the probability $p_{is}$ that click $x_i$ is on subcategory $s$ as
\begin{align}
    p_{is} = \frac{1}{1 + \exp\left(-(\beta_s * v_i + \beta_{t_i,s})\right)}.
\end{align}
In other words, we learn subcategory-specific coefficients $\beta_s$ for being in the vaccine intent window and $\beta_{t,s}$ for each day in the study period.
We compare this model to a nested model where we replace $\beta_{t,s}$ with a single intercept $\beta_0$, so that we no longer control for time.
Across all subcategories, we find that there is no significant change in $\beta_s$ between the two models, so controlling for time does not have a significant impact.
We also find that the learned $\beta_s$'s are very similar to the ratios reported in Table~\ref{tab:near-vi-ratios}, which is expected, since the $\beta_s$'s represent log odds ratios and we report log probability ratios, and odds (i.e., $\frac{p}{1-p}$) and probability are similar when $p$ is small.

We conduct a final supplementary analysis to deepen our understanding of the dynamics of vaccine concerns near vaccine intent.
Previously, we compared click probabilities when a holdout was in their vaccine intent window ($\pm3$ days) versus not in the window. 
Now, instead of only considering a binary variable, we consider a categorical variable $k$ to represent the range of days from $k=-14$ days before vaccine intent to $k=+7$ days after.
For each day in that range, we estimate the probability of clicking on each vaccine subcategory (Figure~\ref{fig:subcat-dynamics}).
We can see that the probability of clicking on every subcategory is elevated close to vaccine intent, but some subcategories are much more elevated than others (e.g., comparing vaccines, Johnson \& Johnson, vaccine incentives), which correspond to the most positive subcategories in Figure~\ref{fig:concerns}d.
We can also see small differences in temporal trends. For example, interest in comparing vaccines and in the Johnson \& Johnson vaccine becomes elevated well before vaccine intent, but interest in normal side effects only grows right before vaccine intent and stays elevated afterwards (perhaps reflecting individuals preparing for their vaccine appointment).
Since all subcategories are elevated near vaccine intent but some more than others, we may want to normalize by the overall increase in vaccine-related interest to tease out differences between subcategories. 
As a secondary analysis, we estimate the probability of clicking on each subcategory per day, \textit{conditioned} on the click being vaccine-related (which is analogous to our analysis in Figure~\ref{fig:concerns}d). 
Now, we can clearly see subcategories that have more elevated interest versus less elevated interest (Figure~\ref{fig:subcat-dynamics-cond}).
Most notably, the bottom row---which represents the most holdout-leaning subcategories from Figure~\ref{fig:concerns}c---exhibits strong drops in relative interest as the individual approaches vaccine intent, especially in seeking exemptions to vaccine requirements, religious concerns about the vaccine, and anti-vaccine messages from high-profile figures and experts.

\begin{figure}
    \centering
    \includegraphics[width=\linewidth]{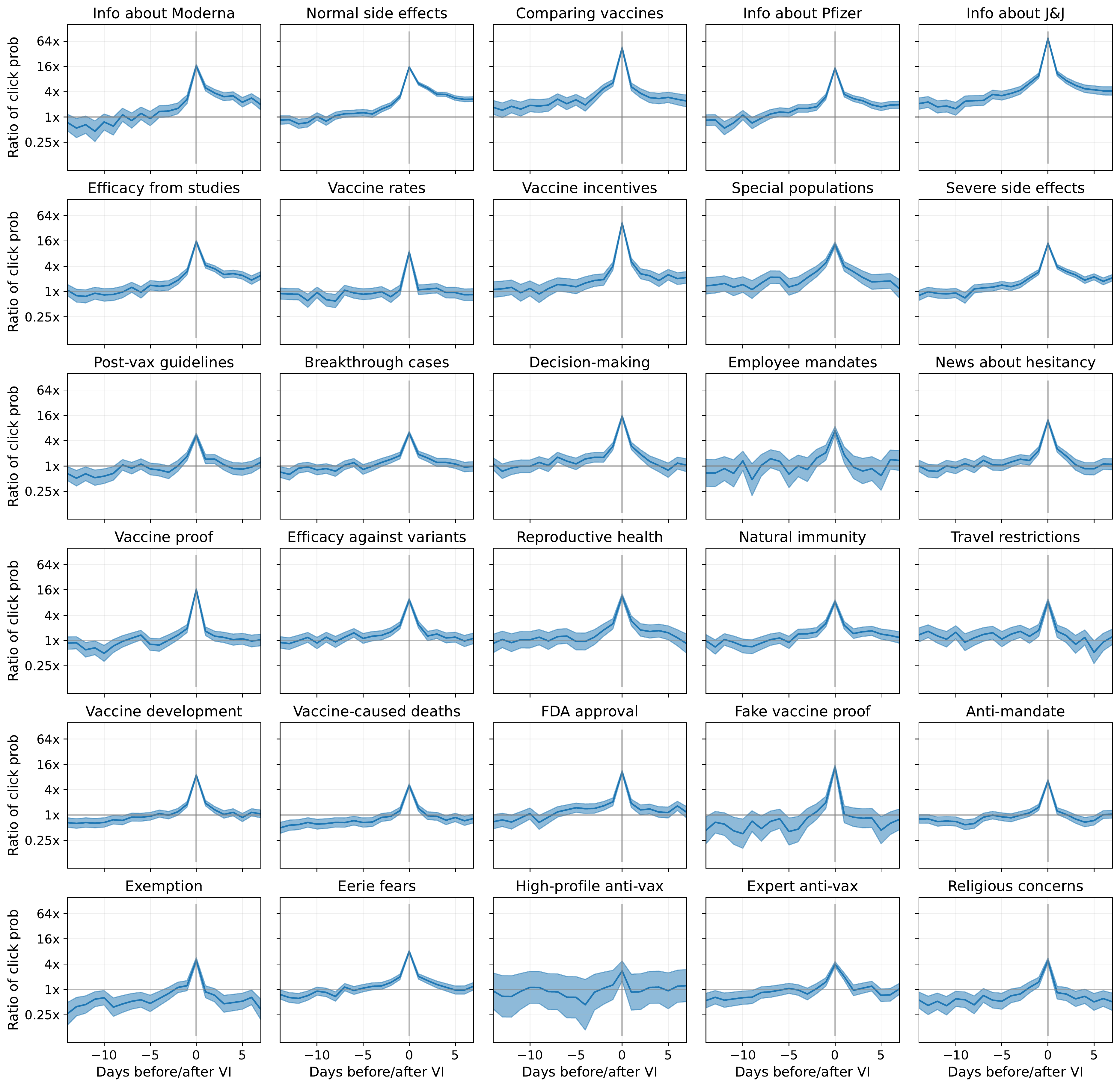}
    \caption{For each day from -14 days before to +7 days after vaccine intent, we measure  holdouts' probability of clicking each subcategory, divided by their baseline probability outside of that range. Interest in all subcategories is elevated near vaccine intent, but certain subcategories are much more elevated than others. Subcategories are ordered from most early adopter-leaning to most holdout-leaning, according to Figure~\ref{fig:concerns}c. Shaded regions represent 95\% CIs.}
    \label{fig:subcat-dynamics}
\end{figure}

\begin{figure}
    \centering
    \includegraphics[width=\linewidth]{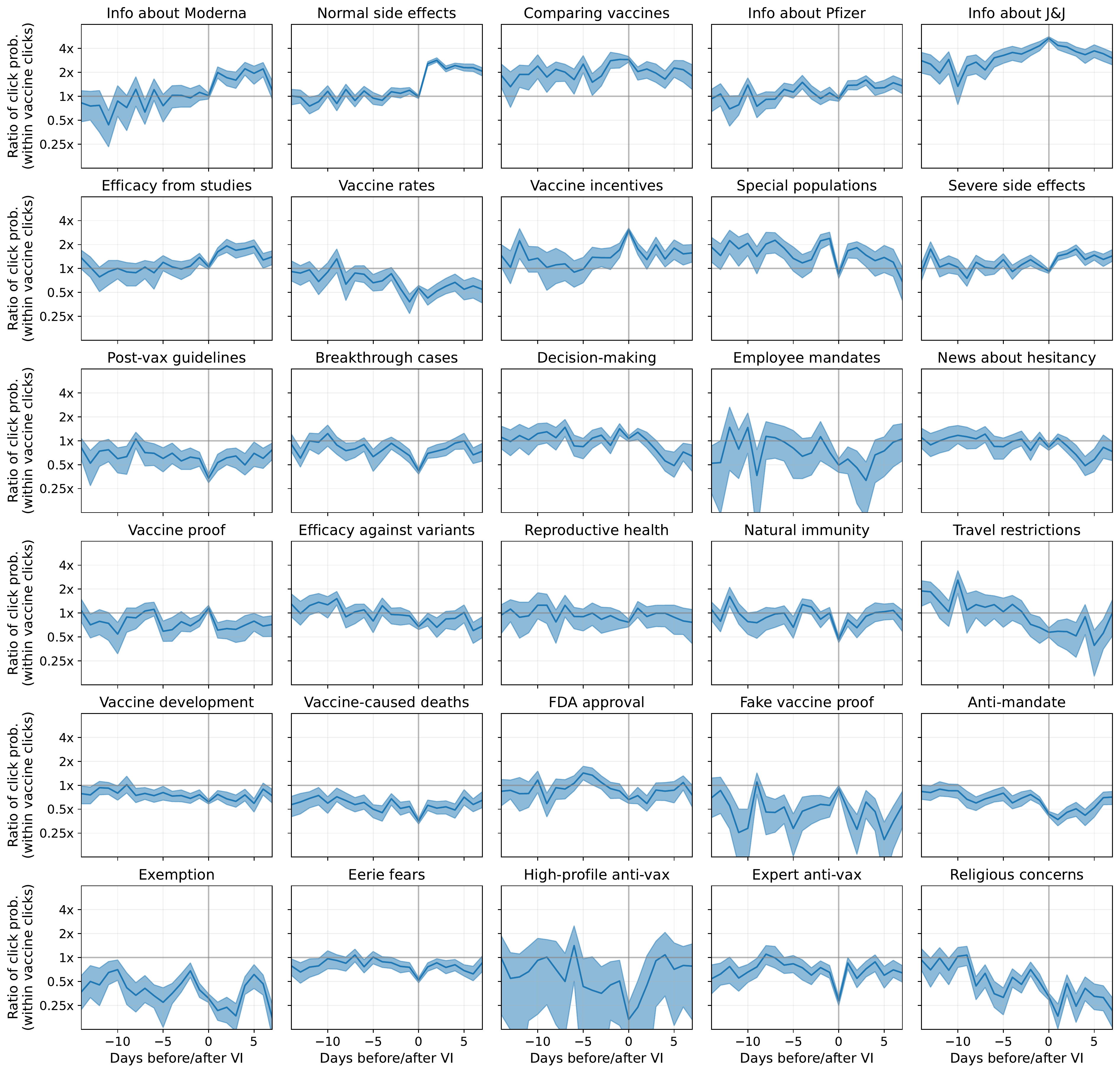}
    \caption{For each day from -14 days before to +7 days after vaccine intent, we measure  holdouts' \textit{relative} probability of clicking each subcategory (conditioned on the click being vaccine-related), divided by their baseline relative probability outside of that range. Measuring within vaccine-related clicks emphasizes differences between subcategories. Subcategories are ordered from most early adopter-leaning to most holdout-leaning, according to Figure~\ref{fig:concerns}c. Shaded regions represent bootstrapped 95\% CIs.}
    \label{fig:subcat-dynamics-cond}
\end{figure}

}
\clearpage
{\customspacing{1}
\section*{References}
\vspace{-5mm}
\bibliographystyle{unsrt}
\bibliography{main}
}
\clearpage
{\customspacing{1}
\paragraph{Acknowledgements.} 
The authors thank 
Ruth Appel, Jonas Barklund, Lisa Cooper, Victor Dibia, Rahul Dodhia, Irena Gao, Kristina Gligorić, Ece Kamar, Paul Koch, Jure Leskovec, Besmira Nushi, Mayana Pereira, Emma Pierson, Yusuf Roohani, Rok Sosic, Albert Sun, Johan Ugander, Michihiro Yasunaga, and members of Johan Ugander's lab for helpful discussions and support.
S.C. conducted this work as an intern at Microsoft.

\paragraph{Author Contributions.} 
S.C. performed computational analysis. All authors jointly analyzed the results and wrote the paper.

\paragraph{Author Information.} 
The authors declare no conflict of interest. Correspondence should be addressed to horvitz@microsoft.com.

\paragraph{Data Availability.} 
Our vaccine intent estimates and ontology of vaccine concerns are publicly available at \github.
Aside from Bing search logs, all of the data sources that we use are publicly available online.
The CDC data, US Census data, and Google search trends can be directly downloaded, and the elections data and Newsguard data can be purchased at the references provided in Section \ref{sec:methods-data}.

\paragraph{Code Availability.} 
Our code for running experiments and generating figures is also publicly available at \github.

}
\end{document}